\documentclass[fleqn,usenatbib]{mnras}

\usepackage{newtxtext,newtxmath}
\usepackage{textgreek}
\usepackage{subcaption}
\usepackage{bm}

\usepackage[T1]{fontenc}

\DeclareRobustCommand{\VAN}[3]{#2}
\let\VANthebibliography\thebibliography
\def\thebibliography{\DeclareRobustCommand{\VAN}[3]{##3}\VANthebibliography}

\usepackage{graphicx}	
\usepackage{amsmath}	

\title[GalactiKit, merger properties with SBI]{GalactiKit: reconstructing mergers from $z=0$ debris using simulation-based inference in Auriga}

\author[A. Sante et al.]{
Andrea Sante,$^{1}$\thanks{E-mail: A.Sante@2022.ljmu.ac.uk}
Daisuke Kawata,$^{2,3}$
Andreea S. Font$^{1}$
and Robert J. J. Grand$^{1}$
\\
$^{1}$Astrophysics Research Institute, Liverpool John Moores University, 146 Brownlow Hill, Liverpool L3 5RF, UK\\
$^{2}$Mullard Space Science Laboratory, University College London, Holmbury St Mary, Dorking, Surrey RH5 6NT, UK\\
$^{3}$National Astronomical Observatory of Japan, 2-21-1 Osawa, Mitaka, Tokyo 181-8588, Japan\\
}

\date{Accepted XXX. Received YYY; in original form ZZZ}

\pubyear{\the\year{}}

\begin{document}
\label{firstpage}
\pagerange{\pageref{firstpage}--\pageref{lastpage}}
\maketitle

\begin{abstract}
We present GalactiKit, a data-driven methodology for estimating the lookback infall time, stellar mass, halo mass and mass ratio of the disrupted progenitors of Milky Way-like galaxies at the time of infall. GalactiKit uses simulation-based inference to extract the information on galaxy formation processes encoded in the Auriga cosmological MHD simulations of Milky Way-mass halos to create a model that relates the properties of mergers to those of the corresponding merger debris at $z=0$. We investigate how well GalactiKit can reconstruct the merger properties given the dynamical, chemical, and the combined chemo-dynamical information of debris.  For this purpose,  three models were implemented considering the following properties of merger debris: (a) total energy and angular momentum, (b) iron-to-hydrogen and alpha-to-iron abundance ratios, and (c) a combination of all of these. We find that the kinematics of the debris can be used to trace the lookback time at which the progenitor was first accreted into the main halo. However, chemical information is necessary for inferring the stellar and halo masses of the progenitors. In both models (b) and (c), the stellar masses are predicted more accurately than the halo masses, which could be related to the scatter in the stellar mass-halo mass relation. Model (c) provides the most accurate predictions for the merger parameters, which suggests that combining chemical and dynamical data of debris can significantly improve the reconstruction of the Milky Way's assembly history. 
\end{abstract}





\section{Introduction}
\label{sec:Intro}

In the standard cosmological model, galaxies form hierarchically by merging with smaller satellite systems over time \citep{WhiteRees_1978}. A natural consequence of this "bottom-up" formation scenario is the assembly of a stellar component in the outer region of galaxies, the stellar halo, dominated by tidally stripped stars once orbiting the cannibalised systems \citep{SearleZinn_1978}. Based on the idea that stellar populations preserve information on the history of the environment since they formed \citep{Eggen_1962, SearleZinn_1978}, stellar halos have become objects of study as a bridge between the present and past of a galaxy's evolution. 

At the first stages of accretion, tidal debris tend to follow the orbit of the merging progenitor as spatially coherent structures, called streams. As the progenitor gradually sinks in the gravitational potential of the host, stellar streams lose their coherence blending with the stars formed \textit{in-situ} \citep{Knebe_2005}. Because galaxies can be regarded, even when merging, as collisionless systems \citep{BinneyTremaine2008}, the probability density of the distribution function of stars around a point in phase-space is conserved over time (Liouville's theorem).  Exploiting this property, \cite{Helmi_1999} devised a methodology to separate phase-mixed structures belonging to different progenitors as overdensities in a new physical space described by the \textit{integrals of motion}, quantities which are constant in time. 

A further method for inferring the assembly history of a galaxy from its present-day stellar content is by "chemical tagging" \citep{FreemanBlandHawthorn_2002}.  Satellite galaxies have an independent chemical evolution history that imprints a unique signature on the chemical composition of forming stars; hence, the abundances of chemical elements can be used to "tag" stars with a similar formation site and time. 

While earlier `archaeological' investigations focused on the stellar halo of the Milky Way \citep[see reviews of][]{Helmi_2020,Deason_2024,Bonaca_streamsrev_2025}, where longer timescales for phase-mixing enable the retention of information about the mergers, more recent studies have extended the search for debris into the disc and inner regions of the Galaxy,  as more data have become available \citep{Arentsen_PIGS_2020, Kawata_MWdisk_2024}. The first hints to the assembly history of the Galaxy were detected with the Sagittarius \citep{Ibata_1994} and Helmi \citep{Helmi_streams_1999} streams, which led to the discovery of examples of late and early accretion events; however, it was only with the first and second data release of the European Space Agency's \textit{Gaia} mission \citep{Gaia_DR1_2016, Gaia_DR2_2018} that a more detailed picture of the past of the Milky Way came into view. The measurements of accurate astrometry and photometry for over a billion stars significantly enhanced the discovery of streams and kinematic substructures associated to merging events. Moreover, the complementary data on radial velocities and chemical abundances for tens of millions of stars obtained with ground-based spectroscopic surveys - such as the Apache Point Observatory Galactic Evolution Experiment (APOGEE, \citealt{Apogee_2017}), the Galactic Archaeology with HERMES (GALAH, \citealt{Galah_2021}) and the Large sky Area Multi-Object fiber Spectroscopic Telescope (LAMOST, \citealt{LAMOST_2012}), to mention a few - pushed further back our gaze into the first stages of the Milky Way evolution \citep{Kruijssen_kraken_2019, Horta_heracles_2021, Belokurov_aurora_2022, Malhan_shiva_2024}. 

The general consensus is that the assembly history of the Milky Way was dominated by a radial, massive accretion event which occurred $\sim$10 billion years ago, named Gaia Sausage/Enceladus (GS/E, \citealt{chiba&beers_2000,brook_2003,Meza_2005,Belokurov_GSE_2018, Helmi_GSE_2018}). This merger is thought to have had a dramatic effect on the proto-Milky Way by dynamically heating the stars in the disc and bringing in a significant amount of gas which led to a burst in star formation \citep{Bignone_2019,Grand_GES_2020,Ciuca_GES_2024}. Although discoveries of a plethora of minor mergers have been claimed from the identification of stellar streams \citep{Malhan_streamfinder_2018} and chemo-dynamical studies in the inner halo \citep{Naidu_substructures_2020,Dodd_catalogue_2023, Horta_substructures_2023}, the GS/E is believed to be the last significant merger, making the assembly history of the Galaxy unusually quiet \citep{Hammer_2007}. Currently, the chronology of the merger events describing the Milky Way assembly history is inferred through the ages of the associated globular cluster and stellar populations. The former are derived empirically from the age-metallicity relation \citep{Kruijssen_kraken_2019, Massari_MWhist_2019}, while the latter can be measured directly with asteroseismology techniques \citep{Miglio_plato_2017, Montalban_2021}, or inferred by comparing photometric and spectroscopic observations to stellar evolution models via isochrone \citep{sandersdas_2018, Queiroz_starhorse_2018} and colour-magnitude diagram fitting \citep{Gallart_cmdfitting_2024}. Similarly, methods for estimating the mass of the disrupted progenitors of the Milky Way involve using empirical relations such as the mass-metallicity relation \citep{Harmsen_massmetallicity_2017, Naidu_massdwarfs_2022}, or fitting the density profile of debris \citep{MackerethBovy_massdebris_2020, Lane_massdebris_2023}.

In this study, we introduce GalactiKit: a data-driven methodology to infer the properties of the disrupted progenitors of the Milky Way. In particular, we focus on retrieving the infall time, stellar mass, halo mass and the halo-mass ratio with the host of mergers at the time of infall. These are determined by performing a Bayesian inference analysis assuming that the progenitor properties are parameters of an undefined model that produces the chemo-dynamical distribution of the merger debris at $z=0$. Despite the model being unknown, there are multiple examples of mergers and phase-space evolution of debris within cosmological simulations; hence, a mapping between the debris and mergers properties can be learned using simulation-based inference (SBI, \citealt{Cranmer_sbi_review_2020}). Using the information contained in the simulations, SBI techniques are designed to perform Bayesian inference on models which are not analytically defined, or which have intractable likelihood distribution functions. Hence, SBI methods can be used to relate observations to the most likely set of parameters that generated them, independently of the degree of knowledge of the underlying model.
Under the assumption that observations can be simulated for a given set of parameters, SBI methods rely on artificial neural networks to approximate the posterior (or likelihood) distributions from the joint distribution of data and parameters. Once either of these quantities are defined, the parameters associated with a given observation can be determined probabilistically within the usual Bayesian formalism by sampling the posterior distribution. 

Previous applications of SBI in Astrophysics include constraining cosmological parameters from galaxy cluster properties \citep{Hahn_simbig_2023, HernandezMartinez_sbicamels_2024} and weak lensing maps \citep{Jeffrey_sbiweallensing_2021}, spectral energy distribution (SED) fitting of galaxies \citep{Hahn_SEDfitting_2022} and estimating compact objects parameters from gravitational-wave data \citep{green_sbiexample_2020}. In the context of estimating the properties of merging galaxies, SBI has recently been used by \cite{Widmark_2025} to develop a framework for inferring the mass density and orbital parameters of tidally perturbed dwarf galaxies based on the observed position and line-of-sight velocity field. 

More closely related to the scope of this work, \cite{Viterbo_CASBI_2024} have developed the Chemical Abundance Simulation Based Inference (CASBI) model, which uses the SBI formalism to predict the stellar mass and infall time of the disrupted components of the stellar halo of Milky Way-mass galaxies based on the oxygen and iron abundances of the debris. Following the idea proposed by \cite{Cunningham_CARDS_2022} that the chemical abundance ratio distributions (CARDs) of stellar halos can be interpreted as the combination of the CARDs of the progenitor galaxies, CASBI was trained and tested on a dataset of artificial halos made up from a catalogue of dwarf galaxies from the NIHAO suite of cosmological simulations \citep{wang_nihao_2015, buck_nihao_2020}. CASBI achieves remarkable accuracy in estimating the stellar massed of the dwarf galaxy components of the artificial halos, but performs less well in terms of estimating their time of infall. 

Here we consider a different approach by developing an SBI pipeline directly on the merger histories of galaxies within the Auriga suite of cosmological simulations \citep{Grand_auriga_2017, Grand_dr_2024}. We also investigate the implications of using the dynamical and/or chemical properties of the debris to inform the inference.

An overview of the Auriga simulations, as well as a description of the procedure to extract and characterize the merger events in the suite, is given in Section \ref{sec:auriga_sims}. A general description of SBI and details about its implementation in the GalactiKit models are included in Section \ref{sec:sbi}. The results of the analysis are presented in Section \ref{sec:results}, while potential applications of the models are proposed in Section \ref{sec:discussions}. We summarise our conclusions in Section \ref{sec:conclusions}. 


\section{The Auriga simulations}
\label{sec:auriga_sims}

The data we use to train Galactikit are taken from the Auriga suite of simulations \citep{Grand_auriga_2017, Grand_dr_2024}. These are cosmological magnetohydrodynamical (MHD) zoom-in simulations of (relatively isolated) Milky Way-mass halos run with the \texttt{AREPO} moving mesh code \citep{Springel_arepo_2010}. 

The halos were selected from the L100N1504 \texttt{EAGLE} dark-matter only simulation \citep{Schaye_eagle_2015}, which was run in a periodic cube of comoving side length of 100 cMpc, with a dark matter particle resolution of $1.15 \times 10^{7} \, M_{\sun}$. The simulation adopts a \textLambda \, cold dark matter (\textLambda CDM) cosmology with model parameters taken from Planck Collaboration XVI (\citeyear{Planck_collab_2014}).

The initial conditions for the re-simulation of a given Auriga halo are prepared at $z=127$. The high-resolution region is defined in the Lagrangian region around the halo, outside of which the resolution drops off with increasing distance. Baryons are added by splitting each dark matter particle into a dark matter particle-gas cell pair with masses given by the cosmological abundance of baryons. The Auriga suite comprises halos simulated primarily at two resolution levels, which are typically referred to as `level 4' and `level 3'. In this work, we consider the `level 4' simulations, which comprise 39 Milky Way analogues in the high (30, $1<M_{200}/[10^{12} \, M_{\sun}]<2$) and low (9, $0.5<M_{200}/[10^{12} \, M_{\sun}]<1$) end of the estimated virial mass range for the Milky Way \citep{Wang_MWmass_2022}. The typical particle mass resolution is $\sim 3 \times 10^{5}$ and $\sim 5 \times 10^{4} \, M_{\sun}$ for dark and baryonic matter, respectively.

The Auriga physics model is described in \cite{Grand_auriga_2017} and briefly summarized herein. Magnetic fields are seeded at the beginning of the simulations in a single direction and with a comoving strength of $10^{-14} \, \mathrm{G}$. Re-ionization is modeled with a spatially uniform, time-dependent ultraviolet background radiation field \citep{Faucher-Giguere_2009} that completes at $z=6$. Primordial and metal-line cooling of gas with self-shielding corrections are also implemented. As gas collapses to densities larger than 0.11 particles per cubic centimetre it enters a sub-grid model for star formation: such gas is modeled as a two-phase interstellar medium, comprised of a cold and hot phase, with an effective equation of state \citep[see][for a detailed description]{Springel_ISM_2003}. Star particles are then created stochastically following a time-dependent, exponential probability function. Each particle represents a single stellar population, with a defined age and metallicity. A Chabrier (\citeyear{Chabrier_imf_2003}) initial mass function is assumed for the stellar evolution of the particle. The chemical enrichment of surrounding gas cells comes from asymptotic giant stars, type-II supernovae,  and type-Ia supernovae. Once halos surpass a mass of $5\times 10^{10} \, M_{\sun}$, black holes are seeded with a mass of $5\times 10^{5} \, M_{\sun}$ at the gravitational potential minimum of the halo, and proceed to grow from gas accretion and merger processes following the prescription in \cite{Springel_bh_2005}. Energetic feedback is modeled phenomenologically as supernovae-driven galactic winds, as well as radiative and thermal energy injection from active galactic nuclei radio and quasar modes. 

The Auriga galaxies have rotationally-supported discs, flat rotation curves and are consistent with the expected stellar mass-halo mass, mass-metallicity and star formation rate relations \citep{Grand_auriga_2017}. Moreover, the Auriga galaxies were directly compared to the observational counterparts from the GHOSTS survey \citep{Radburn-Smith_ghosts_2011, Monachesi_ghosts_2016} by \cite{Monachesi_auriga_2019}, who showed the stellar halos in Auriga reproduce the scatter in stellar mass, surface brightness and metallicity profiles of Milky Way-mass galaxies. However, the Auriga galaxies were also found to have more massive stellar halos due to an extended in-situ component, which is discarded in this analysis.

\subsection{Extracting mergers information}
\label{sec:merger_info}

The $z=0$ stellar composition of the Auriga galaxies is the result of in-situ star formation and merger accretion. For each simulation, the evolutionary histories of all galaxies in the box are encoded in the merger trees, which were computed in post-processing following the \texttt{LHaloTree} algorithm \citep{Springel_2005}. At first, the \texttt{SUBFIND} group finding algorithm \citep{Springel_2001} is used to identify halos and subhalos in all simulation outputs (snapshots). Then, each halo (progenitor) is matched to all the ones having common particles at the subsequent snapshot; a unique descendant is chosen by selecting the halo with the highest number of particles in common weighted by their binding energy. A merger occurs when two or more progenitors point to the same descendant.  

From the merger trees, we can infer the stellar masses well and structural properties of galaxies at any given time.  As part of the public data release of the Auriga Project \citep{Grand_dr_2024}, the accretion history of the star particles within $R_{200}$\footnote{Galactocentric radius enclosing a mass density 200 times higher than the critical density of the universe, often taken as a proxy of the virial radius.} of the main halos, as computed by the merger trees, is reported in the "Accreted particle lists" catalogues.

There are two possible approaches for sorting the accreted particles in their systems of `origin', hence defining the progenitors of the main halo: i)
using the \texttt{RootIndex} list, which associates each particle to the index in the merger tree of the progenitor halo which was bound to at formation; ii) with the \texttt{PeakMassIndex} list, which matches a particle to the merger tree index of the progenitor halo which was bound to at the time the progenitor reached its maximum stellar mass. We decide to consider the second definition because we are interested in retrieving the properties of a progenitor when it first interacts with the main halo, hereby considered as when it first crossed $R_{200}$, which often coincides or shortly follows the time the progenitor undergoes a quenching in star formation \citep{Kawata_2008, Simpson_2018, Font_quenching_2022}.

\begin{figure*}
    \centering
     \begin{subfigure}{0.4\textwidth}
        \centering
        \includegraphics[width=0.8\textwidth]{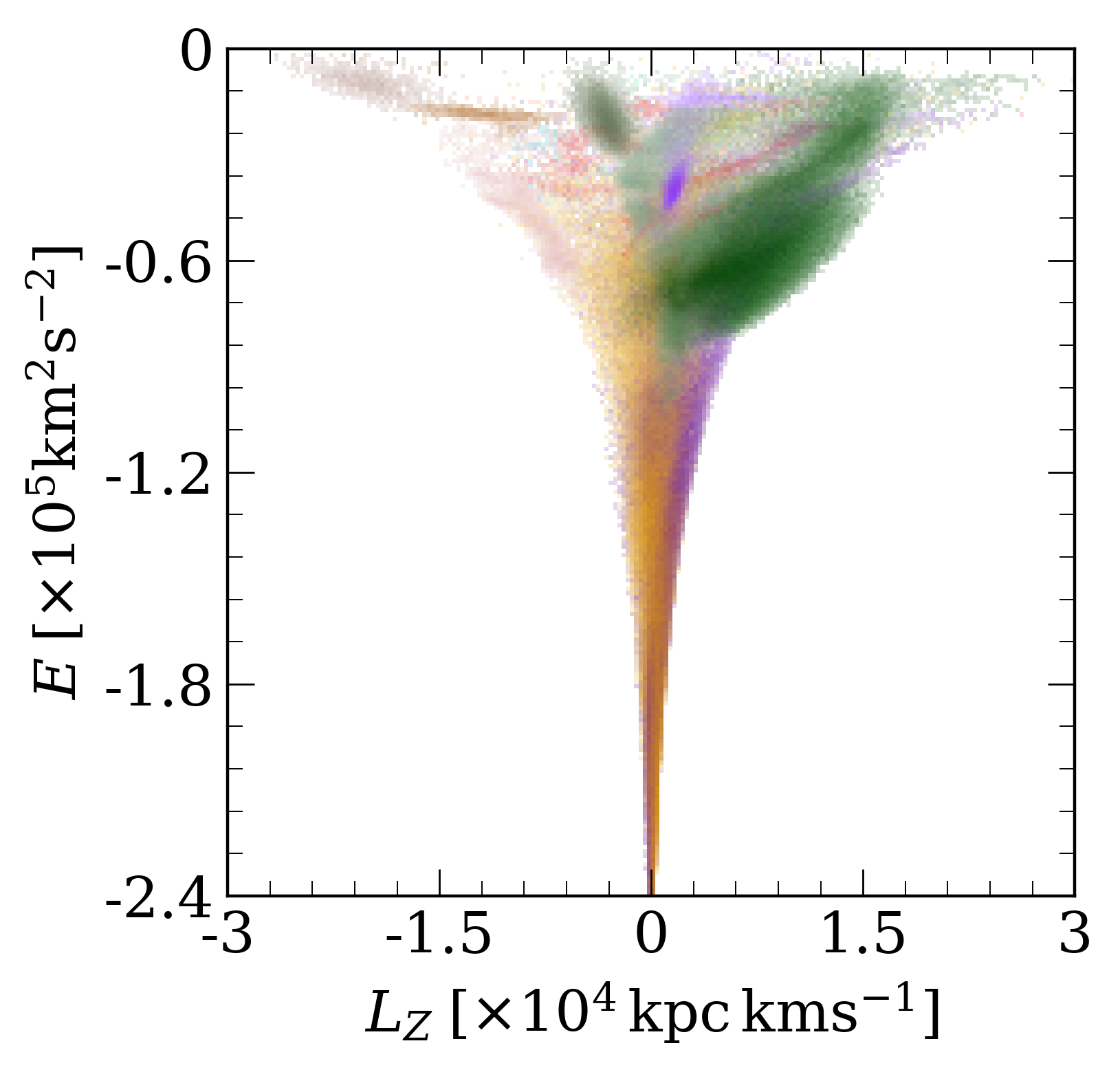}
    \end{subfigure}
    \begin{subfigure}{0.4\textwidth}
        \centering
        \includegraphics[width=0.8\textwidth]{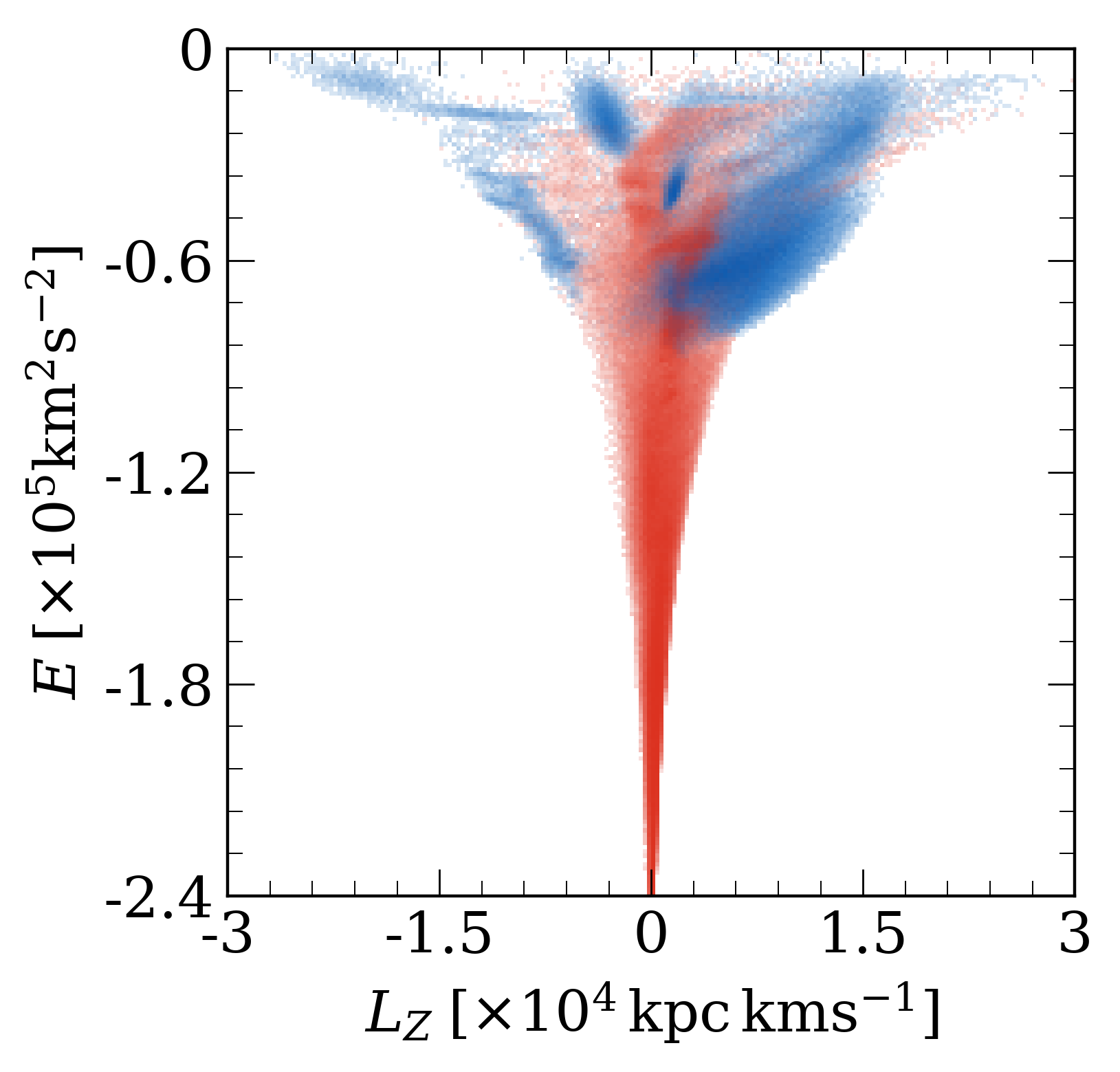}
    \end{subfigure}
    \begin{subfigure}{0.4\textwidth}
        \centering
        \includegraphics[width=0.8\textwidth]{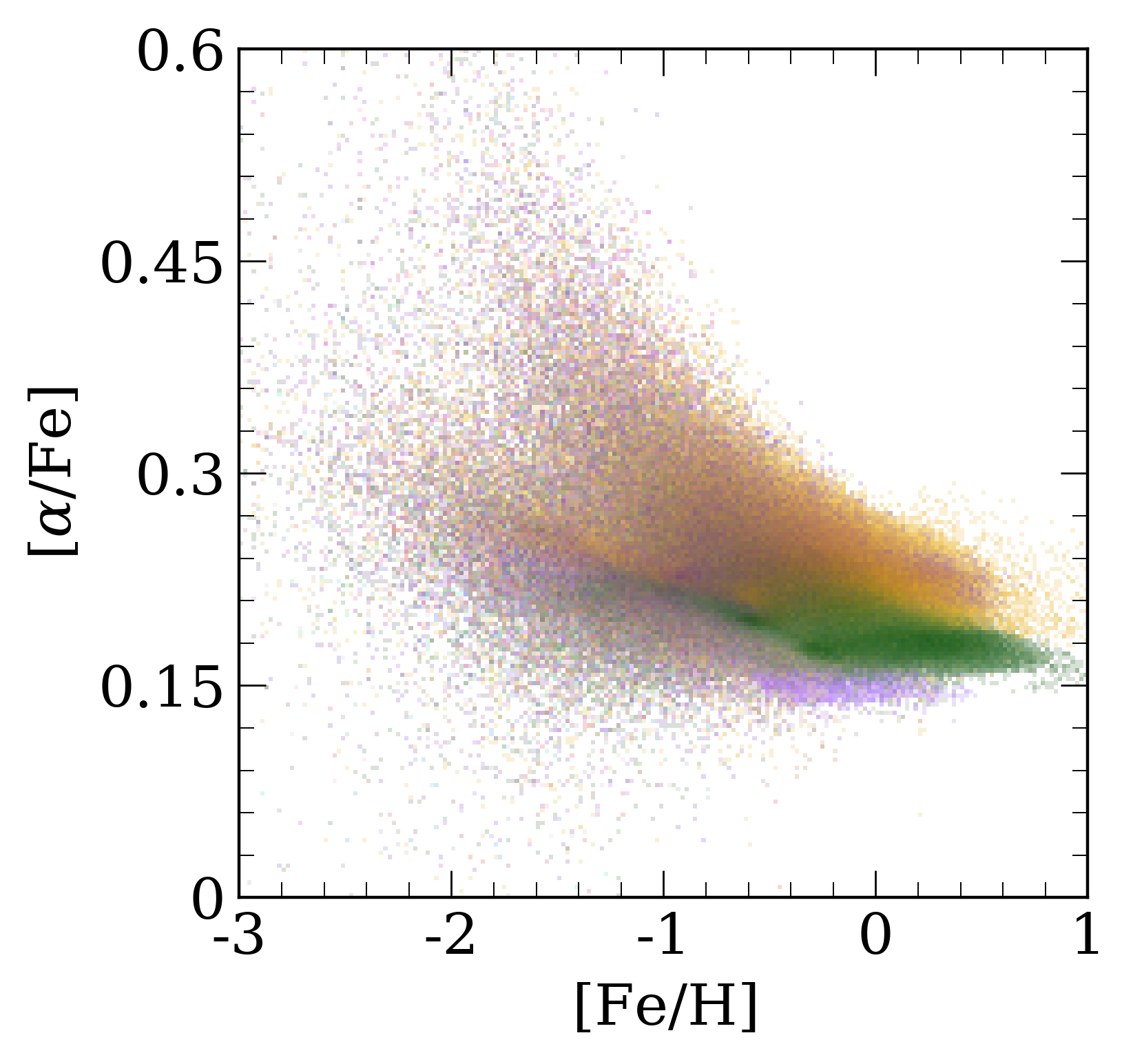}
    \end{subfigure}
    \begin{subfigure}{0.4\textwidth}
        \centering
        \includegraphics[width=0.8\textwidth]{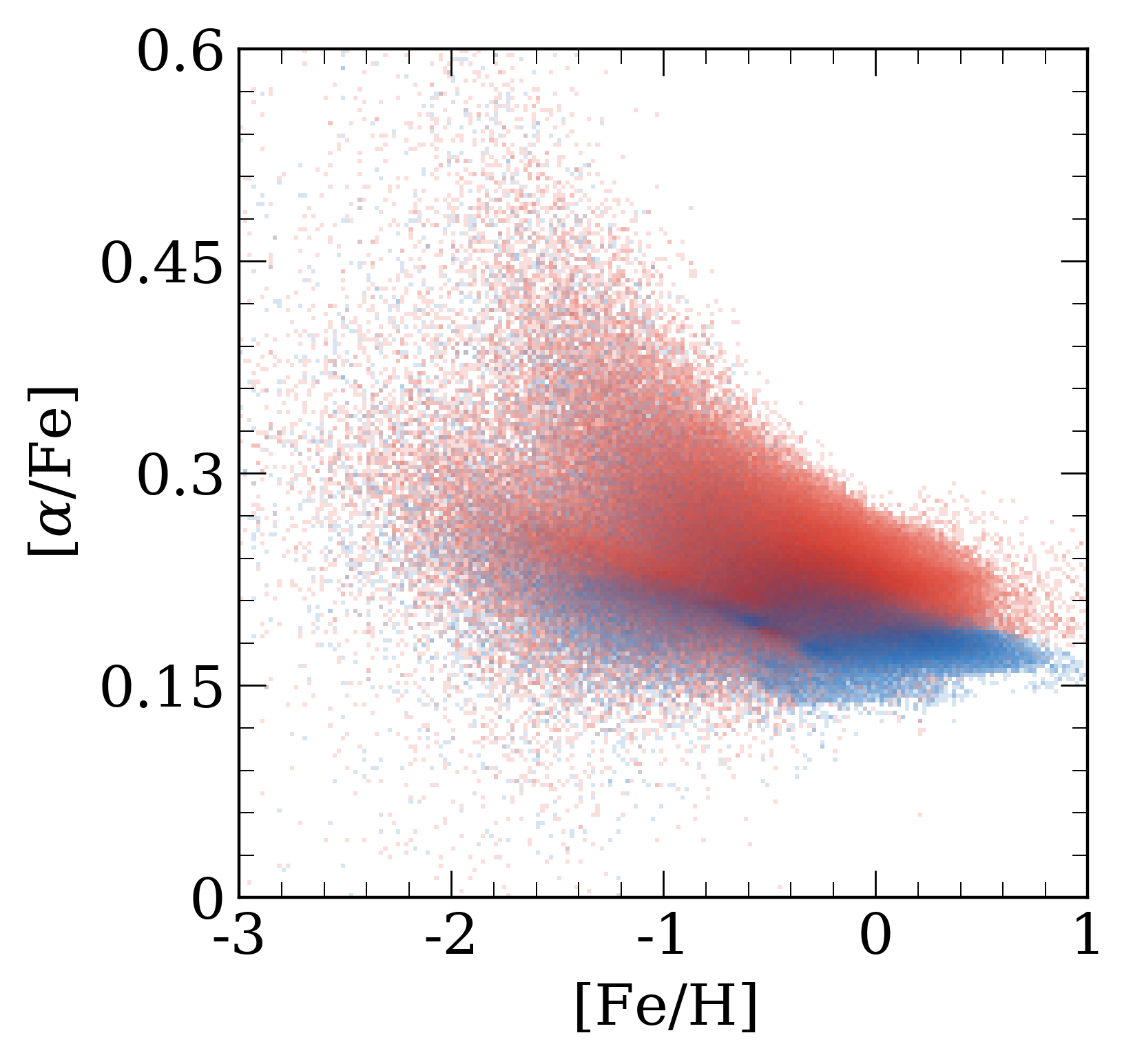}
    \end{subfigure}
    \caption{Accreted star particles located within $R_{200}$ in the Au21 simulation at $z=0$. The top row shows the integrals-of-motion distribution of the particles, while the $[\text{Fe}/\text{H}]-[\alpha/\text{Fe}]$ plane is reported in the bottom. For the plots on the left, the same colour is given to particles coming from the same progenitor galaxy while, in the plots on the right, particles are colour-coded differently if their progenitors are disrupted in the main halo (red) or are orbiting subhalos within $R_{200}$ (blue). }
    \label{fig:Au21}
\end{figure*}

We restrict our investigation to the most significant progenitors in the assembly history of each galaxy by selecting only those contributing with at least 100 star particles. An example of the separation of accreted star particles into the progenitor galaxies in which they formed is reported in Fig. \ref{fig:Au21}, which shows -- for the Au21 simulation -- the $z=0$ distribution of the accreted star particles within $R_{200}$ in the integrals-of-motion (top) and the alpha-iron (bottom) planes. The former is described by the total energy and the $z$-component (perpendicular to the plane of the disc) of angular momentum of particles, while the latter is defined by the iron-to-hydrogen, $[\text{Fe}/\text{H}]$, and alpha-to-iron, $[\alpha/\text{Fe}]$, chemical abundance ratios. In the plots on the left, star particles are colour-coded by the progenitor galaxy in which they formed based on their \texttt{PeakMassIndex} label, while the plots on the right show whether their system of origin is disrupted (red) or appears as an existing satellite (blue) at $z=0$. Although 30 progenitor galaxies were identified for Au21, the majority of accreted stars come from only a few systems, as can be visually distinguished in the plots. Particles from early accretion events -- such as the ones coloured in orange in the left panel of Fig. \ref{fig:Au21} -- tend to have higher abundance of $\alpha$-elements and are more dispersed in the lower region of the integrals-of-motion plane than their later accreted counterparts. 

In the whole suite, a total of 928 progenitors were identified, out of which 792 are taken from the merger trees of the high-mass Auriga simulations ($1<M_{200}/[10^{12} \, M_{\sun}]<2$) and 136 from the low-mass ones ($0.5<M_{200}/[10^{12} \, M_{\sun}]<1$). The majority of them (648) are completely disrupted at $z=0$, and appear as accreted stars;  the rest of stars are from existing satellites (280) orbiting within $R_{200}$.

\section{Simulation-based inference}
\label{sec:sbi}

The underlying idea of the GalactiKit methodology is that some properties of a merging galaxy -- such as the time of infall or the halo mass -- shape the  present-day chemo-dynamical distribution of the corresponding debris. Hence, we assume there is an undefined model $\mathcal{M}$ which, given the vector of properties of a satellite at accretion $\bm{\theta}$, predicts the stellar properties $\bm{x}$ of the corresponding debris at $z=0$. An analytical definition of $\mathcal{M}$ is practically challenging as it should account for complex, coupled physical processes such as the chemical evolution and mass assembly of the host and satellite galaxies in a cosmological context and the phase-space mixing of the accreted debris through dynamical friction within the evolving gravitational potential of the host. However, most of the relevant galaxy formation processes are currently described in the Auriga cosmological simulations at a level of detail which allows for direct comparison between simulated galaxies and their observational counterparts \citep{Monachesi_auriga_2019}. Therefore, simulations can be thought to be already encoding a realistic representation of  $\mathcal{M}$. In order to extract the information encoded in the simulations and make predictions on the properties of a galaxy at infall from the $z=0$ properties of its debris, GalactiKit adopts the SBI framework, which provides a Bayesian description of $\mathcal{M}$. 

In the SBI formalism, given a set of parameters $\bm{\theta}$ and observations $\bm{x}$ from an undefined model or with an intractable likelihood function, if the data generation process can be simulated, the posterior distribution $p(\bm{\theta} \, | \, \bm{x})$ can be evaluated by either: 

\begin{enumerate}
    \item iteratively comparing the observed data to simulations of the model run with different parameters until a similarity threshold is met. This technique is called approximate Bayesian computation \citep[ABC,][]{Ruby_ABC_1984};
    \item approximating the likelihood or posterior distributions with density estimation techniques.
\end{enumerate}
Once the posterior distribution is known, inference on the model parameters for a given observation is performed by sampling the posterior distribution and computing summary statistics.

Because ABC requires simulations to be run at arbitrary points in the $\bm{\theta}$-parameter space and the sample  of merger events $\{\bm{x}, \bm{\theta}\}$ in the Auriga simulations is limited by the assembly history of the simulated halos,  we adopt approach (ii) and implement SBI by developing a deep learning model to approximate the posterior distribution $p(\bm{\theta} \, | \, \bm{x})$.

In this analysis, we focus on the following properties $\bm{\theta}$ of the progenitor galaxies:

\begin{itemize}
    \item infall time ($\tau$), defined as the lookback time at which the satellite first crossed $R_{200}$;
    \item stellar mass and halo mass ($M_{*}$ and $M$, respectively) at infall;
    \item the merger mass ratio (MMR), defined as the ratio between the halo mass of the satellite and that of the main galaxy at infall. 
\end{itemize}

To ensure a strong correlation between $\bm{x}$ and $\bm{\theta}$, we consider the debris properties which are known to be physically related to the merger variables: 

\begin{itemize}
    \item total energy ($E$), which is related to the infall time through the infall time-binding energy relation of subhalos and debris \citep{Rocha_2012, GarciaBethencourt_2023};
    \item total specific angular momentum ($L$) which, like the total energy or the radial action, is an integral of motion for a star in a spherical potential;
    \item $[\text{Fe}/\text{H}]$, which, as a proxy of metallicity, is correlated to the mass of a galaxy through the mass-metallicity relation \citep{Harmsen_massmetallicity_2017}.
    \item $[\alpha/\text{Fe}]$, which relates to the infall time of the progenitor as stars from early accreted galaxies  have, in general, higher abundance of $\alpha$-elements compared to stars from late mergers \citep{Font_2006}.
\end{itemize}

The posterior estimation is performed with the Masked Autoregressive Flow (MAF, \citealt{Papamakarios_MAF_2017}) technique. In brief, MAFs are generative models that approximate complex probability distributions defined by the training data. By drawing a random number from a normal distribution, MAFs generate a new data point through a series of parameterized bijective transformations; this is equivalent to sampling from the approximated probability distribution of the training data. The models are trained by tuning the parameters of the transformations such that the negative log-probability of the training data is minimized. 

A more detailed mathematical description of the SBI framework and MAFs is reported in Section \ref{sec:maths}, while details on the model implementation and training are presented in Section \ref{sec:data_preprocessing}. The reader who is not interested in the detailed methodology can skip these sections and refer directly to the results of the analysis in Section \ref{sec:results}.

\begin{figure*}
    \centering
    \includegraphics[width=0.5\textwidth]{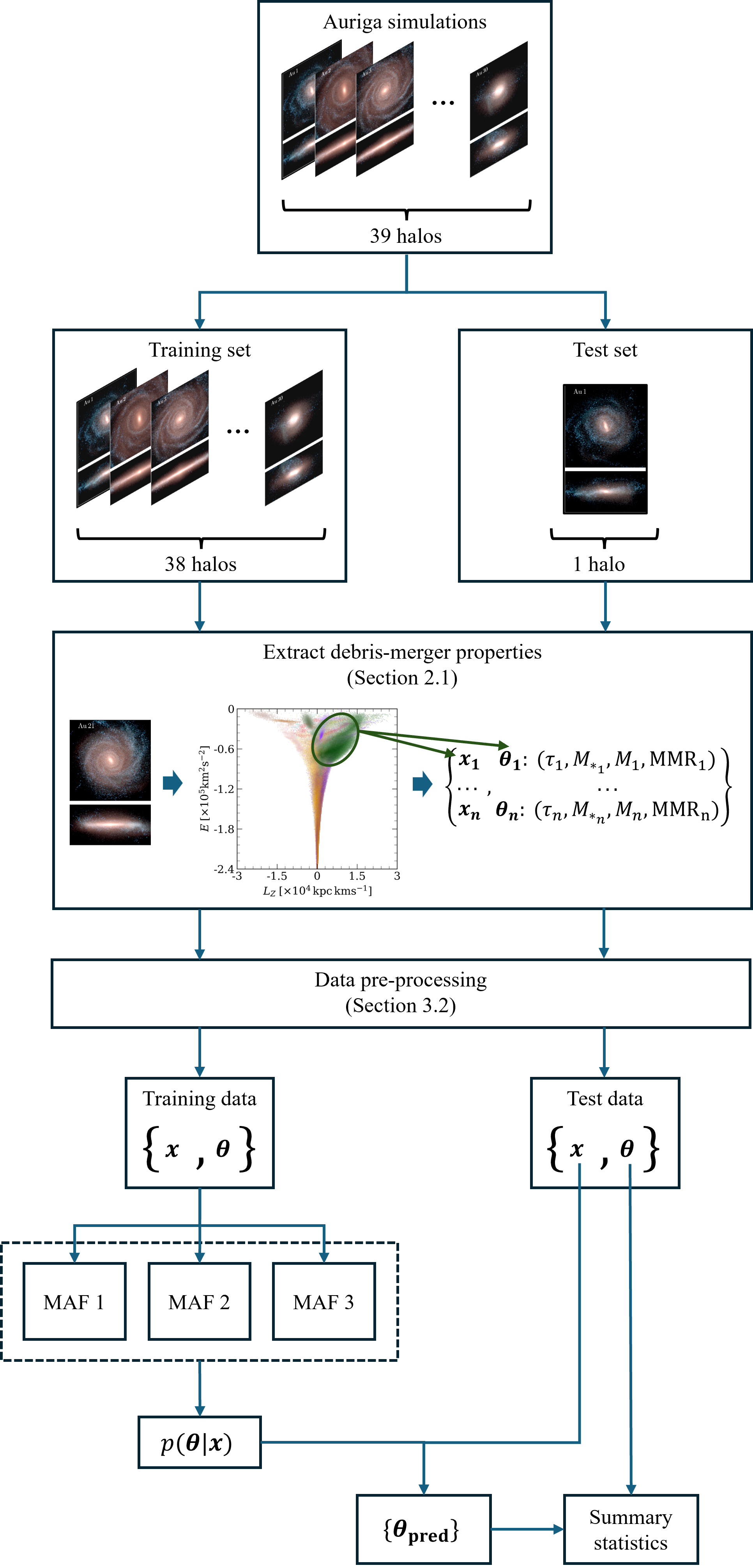}
    \caption{Overview of the GalactiKit methodology. The Auriga simulations were first split into a training and test sets. For both cases, the accreted stars within $R_{200}$ of the main halo were associated to the corresponding progenitor galaxy creating a set of merger-debris pairs $\{\bm{x},\bm{\theta}\}$; this was then pre-processed to be used as input to the MAF models approximating $p(\bm{\theta} \,| \, \bm{x})$. The quality of the estimation was assessed through summary statistics comparing the inferred values to the actual ones for the properties of mergers in the test. The images of the Auriga galaxies are taken from \citet{Grand_auriga_2017}.}
    \label{fig:flowchart}
\end{figure*}

An overview of the GalactiKit methodology is shown in Fig. \ref{fig:flowchart}. Initially, the simulations in the Auriga suite were separated into a training set, used to train the MAF models estimating $p(\bm{\theta} \, | \, \bm{x})$, and a test set, used for validation. For simulations in both sets, the information from the merger tree was used to define a set of merger-debris pairs $\{\bm{x},\bm{\theta}\}$ by grouping the accreted stars orbiting within $R_{200}$ of the main halo at $z=0$ into the corresponding progenitor galaxies as defined in Section \ref{sec:merger_info}. The resulting training and test datasets were then processed (see Section \ref{sec:data_preprocessing}) to become valid inputs for the MAF models. While multiple samples were drawn for the same merger from the debris distribution in the training galaxies depending on the total number of stars found in $R_{200}$ at $z=0$, only one sample per merger was drawn in the test galaxy to avoid results being biased towards major mergers. After training, the quality of the estimation of $p(\bm{\theta} \, | \, \bm{x})$ was determined by comparing the actual properties of the mergers in the test galaxy with the ones inferred by sampling the MAF models. To obtain a statistical representation, 1,000 $\bm{\theta}$-samples were generated per merger using the debris data as input to the models. In order to cross-validate our results, the 39 halos were excluded one at a time from the training set and used for testing. 

\subsection{Density estimation with Masked Autoregressive Flows}
\label{sec:maths}

Consider the model $\mathcal{M}$ relating the properties of a satellite galaxy at the time of infall ($\bm{\theta}$) with the $z=0$ chemo-dynamical properties of the corresponding debris ($\bm{x}$).
Applying Bayes' theorem,  the degree of belief that a certain $\bm{\theta}$ leads to a realisation $\bm{x}$ of the model can be measured by evaluating the posterior probability: 

\begin{equation}
    p(\bm{\theta} \, | \, \bm{x} ) = \frac{p(\bm{x} \, | \, \bm{\theta} ) \; p(\bm{\theta} )}{  \int p(\bm{x} \, | \, \bm{\theta} ) \, d\theta },
\end{equation}
where $p(\bm{\theta})$ is the prior, encoding the data-independent information on the possible merger parameters, and $p(\bm{x} \, | \, \bm{\theta} )$ is the likelihood, measuring the probability of observing a certain configuration of the merger debris distribution given certain merger parameters. Hence, for an observed distribution of merger debris $\bm{x_{\mathrm{obs}}}$, the most likely set of properties of the corresponding progenitor galaxy $\bm{\theta}$ can be inferred by sampling the posterior probability $p(\bm{\theta} \, | \, \bm{x_{\mathrm{obs}}} )$. However, as the likelihood is unknown, the posterior cannot be evaluated explicitly.  This problem can be overcome adopting the SBI framework, where simulations of the model $\mathcal{M}$ are used to obtain an approximation of the posterior distribution. In particular, the posterior distribution is modelled as a distribution function whose parameter values depends on the vector of debris properties $\bm{x_{\mathrm{obs}}}$ used for conditioning the posterior, i.e.

\begin{equation}
    p(\bm{\theta} \, | \, \bm{x_{\mathrm{obs}}} )\approx p_{\bm{\phi}}(\bm{\theta}\, |\, \bm{x_{\mathrm{obs}}}),  
\label{eq:variational}
\end{equation}
where $\bm{\phi}=h(\bm{x_{\mathrm{obs}}})$ is the vector of parameters of the \textit{variational} distribution $p_{\bm{\phi}}$, and the function $h$ represents the mapping between the observations of the model $\bm{x}$ and the $\bm{\phi}$ parameters. Because only the internal parameters change depending on the conditioning input, $p_{\bm{\phi}}$ is the same distribution function for any value of $\bm{x_{\mathrm{obs}}}$; hence, the posterior distribution is \textit{amortised}, i.e. once the mapping between distribution parameters and data is defined, inference on the merger properties $\bm{\theta}$ can be performed for any value of the debris properties $\bm{x}$. 

In order to implement the posterior approximation described in equation (\ref{eq:variational}), we use Masked Autoregressive Flow (MAF), a neural network technique which is based on the idea that generating data with an autoregressive model is equivalent to a normalizing flow \citep{Kingma2016}. A brief description of the mathematical formalism of MAF applied to the inference of merger parameters is presented below; we refer to \cite{Papamakarios_MAF_2017} for details on the algorithm. 

Autoregressive models \citep{URIA_autoregressive_2016} are artificial neural networks used for density estimation, and are based on the idea that any distribution in $D$ dimensions can be represented using the probability product rule as a combination of the conditional probabilities in the single dimensions. Hence, the posterior distribution $p(\bm{\theta} \, | \, \bm{x} )$ of a merger defined by a galaxy whose properties are $\bm{\theta}=(\tau, M_*, M,\mathrm{MMR})$ and debris with a certain vector of properties $\bm{x}$ at $z=0$, can be defined as:

\begin{equation}
    p(\bm{\theta} \, | \, \bm{x} ) = \bm{\prod}_{d=1}^{D=4} p(\theta_{d} \, | \, \bm{\theta}_{<d} , \bm{x}),
\end{equation}
where $\bm{\theta}_{<d} = (\theta_1, ..., \theta_{d-1})$ is a vector whose components are the first $d-1$ components of the $4$-dimensional merger parameter vector $\bm{\theta}$. 

In MAF, the conditional probabilities are modelled as single Gaussians whose parameters are computed using Masked Autoencoders for Distribution Estimation (MADE, \citealt{Germain_MADE_2015}) models:

\begin{align}
    p(\theta_{d} \, | \, \bm{\theta}_{<d}, \bm{x} ) \approx p_{\phi_d}(\theta_d\, | \bm{\theta}_{<d},\bm{x})=  \mathcal{N} \left( \theta_{d} \, | \, \mu_{d}, (\text{exp} \, \alpha_{d})^2 \right),
\end{align}
with $\mu_{d}= h_{\mu_{d}} (\bm{\theta}_{<d},\bm{x})$ and $\alpha_{d} = h_{\alpha_{d}} (\bm{\theta}_{<d}, \bm{x})$, where $h_{\mu_{d}}$ and $h_{\alpha_{d}}$ represent the MADE outputs as the mean and log-standard deviation of the $d^{\text{th}}$ conditional probability. Hence, the $p(\theta_{d} \, | \, \bm{\theta}_{<d}, \bm{x} )$ conditionals are approximated by Gaussian variational distributions whose parameters, $\phi_d=\{\mu_d, \alpha_d\}$, are computed from an artificial neural network which takes as input the $d-1$ previous components of $\bm{\theta}$ and the conditions $\bm{x}$. In this framework, sampling is equivalent to generating a new point in the $\bm{\theta}$-parameter space by explicitly computing each dimension as 

\begin{equation}
    \theta_{d} = u_d \, \text{exp} \, \alpha_{d} + \mu_{d},
\label{eq:transformation}
\end{equation}
where $u_{d} \sim \mathcal{N}(0,1)$. Hence, equivalently to a normalizing flow \citep{Papamakarios_NF_2021}, the model is mapping a point $\bm{u}=(u_1,...,u_d)$ in the base distribution to the merger parameter space through an invertible and differentiable transformation: $\bm{\theta} = f(\bm{u})$. Hence, the variational distribution approximating the posterior can be expressed in terms of the base distribution $p_{u}$ as 

\begin{equation}
    p_{\bm{\phi}}(\bm{\theta} \, | \, \bm{x} ) = p_{u}(f^{-1}(\bm{\theta})) \, \left | \frac{\partial f^{-1}}{\partial \bm{\theta} }\right |,
\label{eq:flow}
\end{equation}
where $\left | \frac{\partial f^{-1}}{\partial \bm{\theta} }\right |$ is the determinant of the Jacobian matrix of $f^{-1}$, which for the transformation defined in equation (\ref{eq:transformation}) is $\left | \frac{\partial f^{-1}}{\partial \bm{\theta} }\right | = \exp\left(-\sum_{d}^{D=4}\alpha_{d}\right)$.
Considering the model as a normalizing flow, it is possible to increase the expressivity of MAF by stacking multiple instances as a sequence of transformations. Thus, a sample of merger parameters generated by a MAF model with $N$ transformations is computed as

\begin{equation}
\label{eq:sampling}
\bm{\theta} = \mathcal{F}(\bm{u}), \; \text{with} \; \mathcal{F}= f_{1} \circ ... \circ f_{N} \; \text{and} \; \bm{u} \sim \mathcal{N}(0, \bm{\text{I}}),
\end{equation}
where $\bm{\text{I}}$ is the identity matrix and each $f$ represents a single transformation, which includes an ensemble of Gaussian conditionals and the MADE model, which take the vector of debris properties $\bm{x}$ and the relevant components of $\bm{\theta}$ as inputs and return the corresponding mean and log-standard deviation. Because a combinations of invertible functions is also invertible and the overall Jacobian determinant is the product of the Jacobian determinants of the single transformations, equation (\ref{eq:flow}) can be adapted to the case of an arbitrary  number of transformation defined in equation (\ref{eq:sampling}), and written in log-form as 

\begin{equation}
    \log\,p_{\bm{\phi}}(\bm{\theta} \, | \, \bm{x} ) = \log\,p_{u}(\bm{u}) \, + \, \sum_{i=1}^{N}\,\log\,\left | \frac{\partial f_{i}^{-1}}{\partial \bm{\theta}_i }\right |,
\end{equation}
where $\bm{\theta}_i=f_i\,\circ ... \circ \, f_1(\bm{u})$ is an intermediate state of the flow $\mathcal{F}$ relating the base distribution to the one of the merger properties $\bm{\theta}$.
During the training routine, the internal parameters of the MADE models, which are used to compute the vectors $\bm{\mu_i}$ and $\bm{\alpha_{i}}$ defining each transformation $f_{i}$, are updated in order to maximise the total log likelihood $\sum_{n}\log\,p_{\bm{\phi}}(\bm{\theta}_n\,|\,\bm{x}_n)$ of the training data,  where $\bm{x}_n,\bm{\theta}_n \, \sim \, p(\bm{\theta},\bm{x})$, which is the joint probability distribution of merger and debris properties defined by the progenitor galaxies in the Auriga simulations.

Once the model has been trained, a new sample of the posterior distribution of the progenitor parameters given a certain observation of debris properties at $z=0$ can be obtained by drawing a random number from a Gaussian distribution and applying the series of data-dependent transformations specified by the MADE models as described in equation (\ref{eq:sampling}).

\subsection{Data pre-processing and training implementation}
\label{sec:data_preprocessing}

Before being used to train the MAF model, the progenitors $\bm{\theta}$ and merger debris $\bm{x}$ properties were processed as follows:

\begin{enumerate}

    \item The distributions of the debris in $E$ and $L$ were transformed by taking the square root of the corresponding absolute values, to compensate for the skewness of the original distributions.
    We also considered scaling the distributions of the debris based on the structural, present-day properties of the galaxies; namely, the maximum rotational velocity, scale radius, $R_{200}$ and $M_{200}$. However, we find that this leads to little or no improvements in the posterior estimation.

    \item Outliers from the interquartile (i.e. $25^{\rm th}$ and $75^{\rm th}$ percentile) ranges of the debris distributions in $E$, $L$, $[\text{Fe}/\text{H}]$, and $[\alpha/\text{Fe}]$ were detected and discarded. This is performed in preparation of the next step, as particles whose properties lie at the margin of the distributions might skew the training data while not representing the majority of the debris from a given merger. After this procedure, all merger events that fell below the 100 star threshold were also removed from the analysis. 
    To ensure all inputs of the models are on the same scale, the resulting distributions were standardized by subtracting the mean and dividing by the standard deviation in each dimension.

    \item Because the number of particles representing the merger debris depends on the progenitor mass, while MAF requires an input of fixed dimensionality, $N$ samples of 100 stars were drawn from each debris distribution. In order to account for the intrinsic variability in $\bm{x}$ when sampling large progenitors, the number of samples for a given merger was calculated in terms of the number of associated accreted stars $S$, as $N=\mathrm{min}(10, \, \left \lfloor{S/100}\right \rfloor )$.

    \item To provide a balanced representation of all mergers in the training set, the debris distributions of mergers with uncommon parameters were oversampled. For this purpose, a k-means model was applied to the $\bm{\theta}$-parameter space to form 20 groups with similar properties. Then, the number of samples from mergers in each group was counted. By imposing that each group has to contribute the same number of samples, a balanced data set was achieved by drawing more samples from the debris of the mergers from underrepresented groups, with each contributing with an equal number of extra samples.
    
\end{enumerate}

The above process returns a dataset composed of merger-debris pairs which more uniformly covers the $\bm{\theta}$-parameter space, is normally distributed in $\bm{x}$, and has homogeneous samples (i.e. all sharing the same dimensionality $D\times100$, where $D$ is the number of stellar properties considered).

The posterior distribution $p(\bm{\theta} \, | \, \bm{x})$ was modeled with an ensemble of three MAFs to avoid overfitting, which is more likely to affect a single density estimator. All the MAF models share the same initial configuration (i.e. number of transformations, number of neurons in the hidden layers, etc.), but are trained separately; because the optimization of the internal parameters of an artificial neural network is a stochastic process, the resulting three MAFs are effectively three different models of the posterior distribution. The implementation of the SBI framework was performed using the Learning the Universe Implicit Likelihood Inference (LtU-ILI) pipeline \citep{Ho_LTUILI_2024} with the \texttt{sbi} backend. The number of transformations in each MAF, the length of the hidden layers in the MADE models, and the training parameters (batch size and learning rate) were derived using the \texttt{Optuna} hyperparameter optimisation framework \citep{akiba_Optuna_2019} setting the log-likelihood of a fixed validation data set as objective function. The resulting MAF models used in the analysis comprised 22 MADE layers with 463 neurons per hidden layer and were trained on batches of 2569 examples with a learning rate of 0.001.


\section{Results}
\label{sec:results}

\begin{figure*}
    \centering
    \begin{subfigure}{\textwidth}
        \centering
        \includegraphics[width=0.8\textwidth]{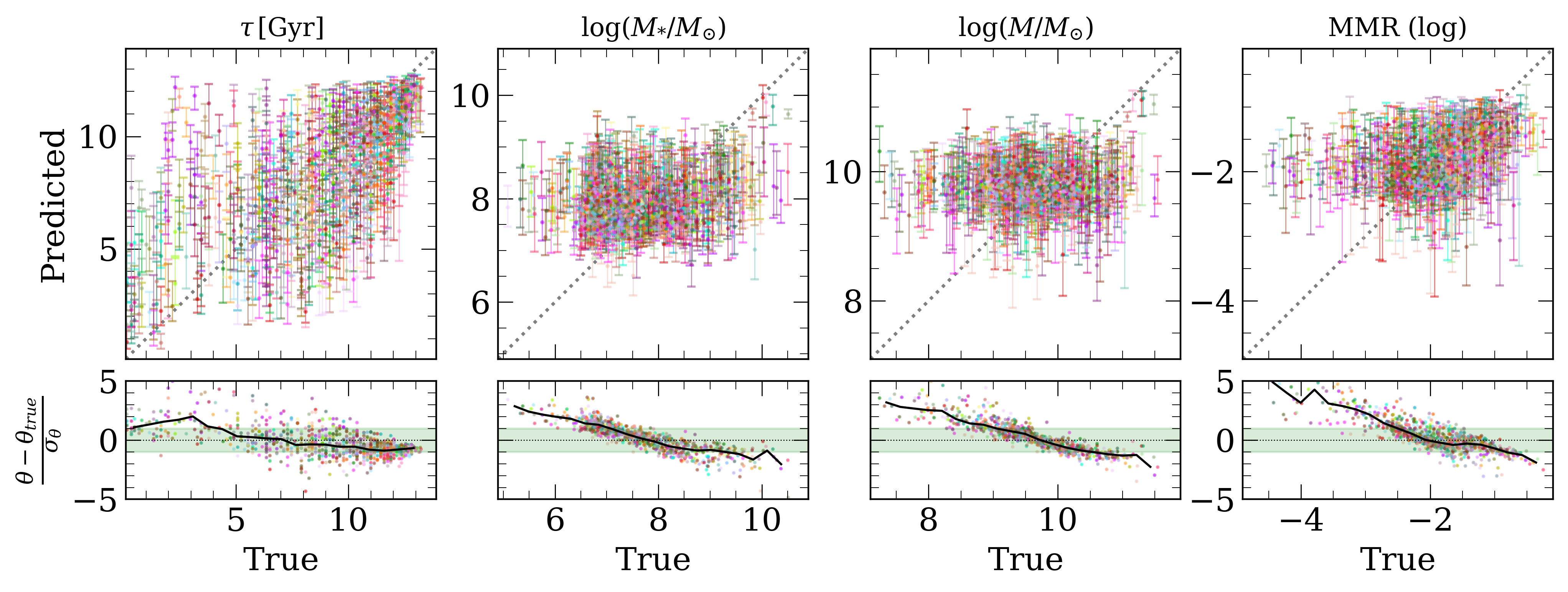}
        \caption{}
        \label{fig:subfig1}
    \end{subfigure}
    \begin{subfigure}{\textwidth}
        \centering
        \includegraphics[width=0.8\textwidth]{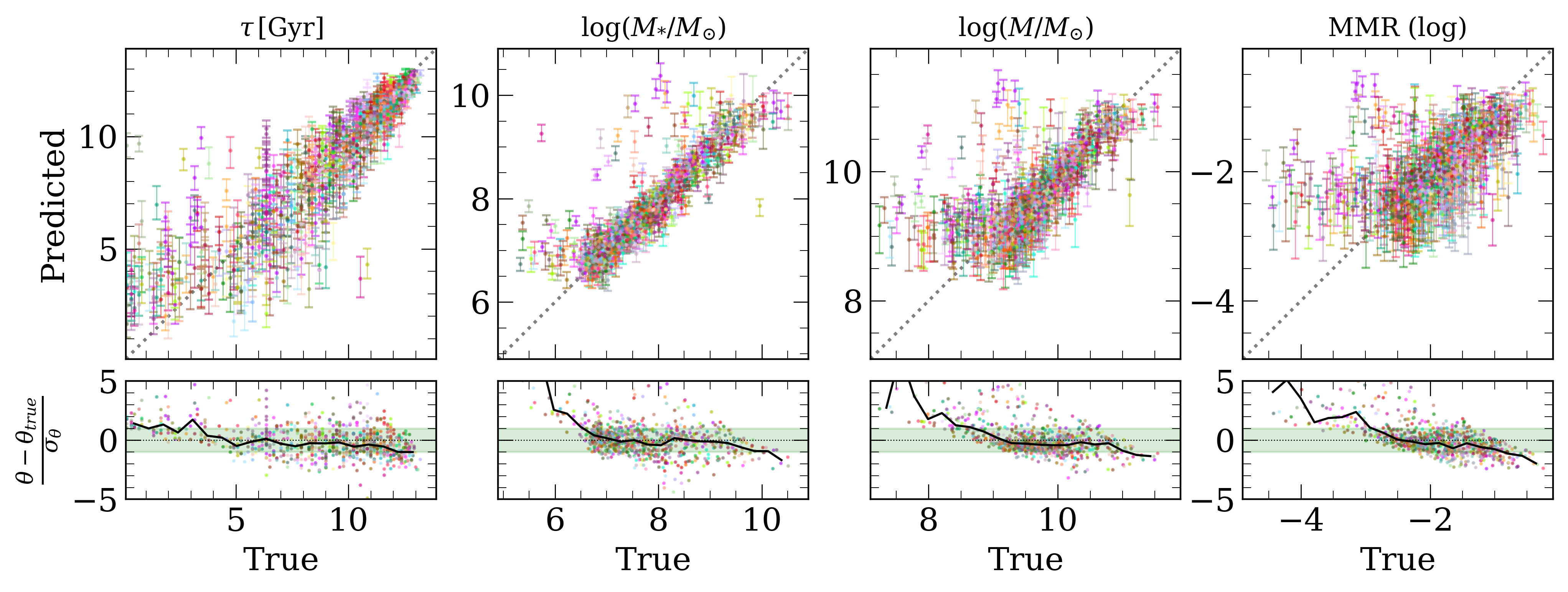}
        \caption{}
        \label{fig:subfig2}
    \end{subfigure}
    \begin{subfigure}{\textwidth}
        \centering
        \includegraphics[width=0.8\textwidth]{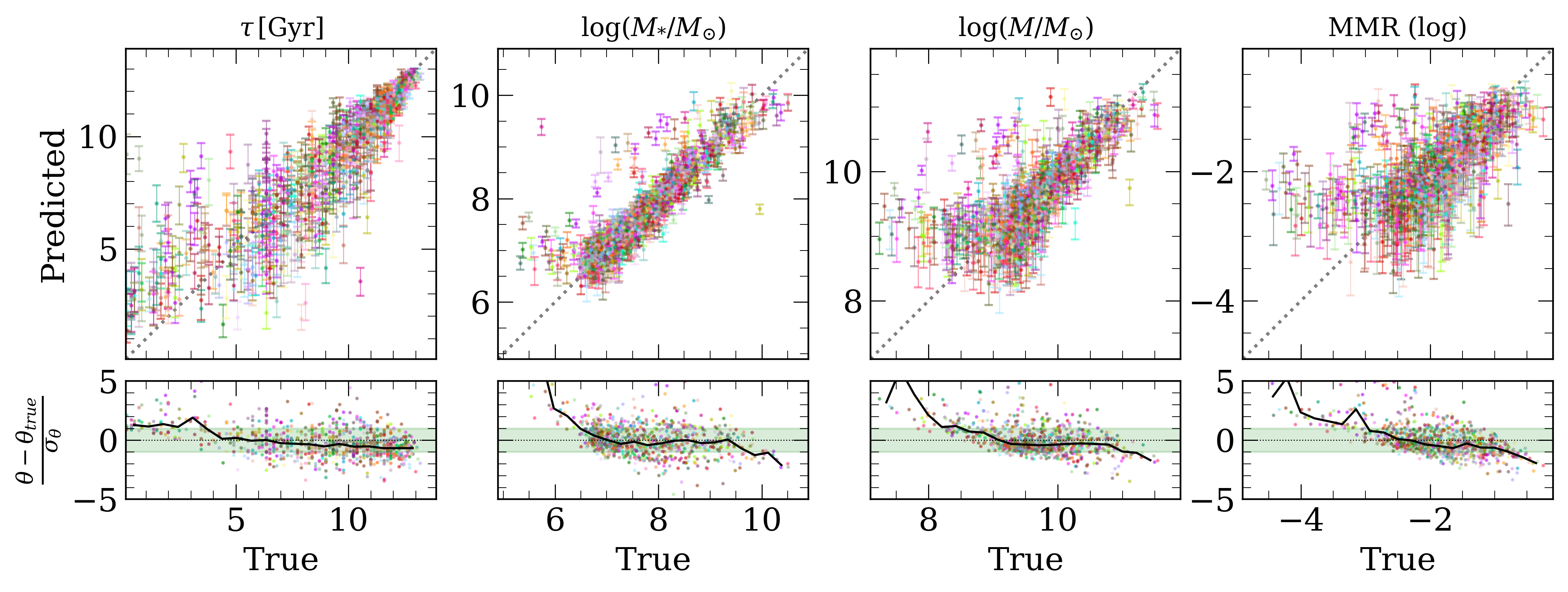}
        \caption{}
        \label{fig:subfig3}
    \end{subfigure}
    \caption{Merger parameters inferred by SBI models trained on different combinations of $z=0$ debris properties: (a) $E$, $L$ (top), (b) $[\mathrm{Fe}/\mathrm{H}]$, $[\alpha/\mathrm{Fe}]$ (middle) and (c) $E$, $L$, $[\mathrm{Fe}/\mathrm{H}]$ and $[\alpha/\mathrm{Fe}]$ (bottom). Each point represents the median value of the parameter samples, while the error bar shows the extent of the $34^{\rm th}-68^{\rm th}$ percentile range of the distribution. A unique colour is associated to merger events associated to a specific galaxy, for which case the SBI models used for inference were trained on the mergers from the remaining galaxies in the suite.}
    \label{fig:cross_validation_cc_galaxy}
\end{figure*}

Three implementation of GalactiKit were developed to estimate the posterior distribution of the model $\mathcal{M}$ described in Section \ref{sec:sbi} for different combinations of debris properties $\bm{x}$, namely: (a) $E, \,L$; (b) $[\text{Fe}/\text{H}], \, [\alpha/\text{Fe}]$; (c) $E, \, L, \, [\text{Fe}/\text{H}], \, [\alpha/\text{Fe}]$. The target merger parameters  were fixed to $\bm{\theta} = (\tau, \, \mathrm{log}(M_{*}/M_{\sun}), \, \mathrm{log}(M/M_{\sun}), \, \mathrm{log}(\text{MMR}))$ for all cases. The aim was to investigate to what extent the properties of a merger can be reconstructed when different degrees of debris information are available. 
For each combination of $\bm{x}$, the results of the cross-validation procedure are reported in Fig. \ref{fig:cross_validation_cc_galaxy} as the inferred against the actual parameters for the mergers in the test galaxies. In this paper, `true' indicates the parameter values as directly measured in the Auriga simulations, while `predicted' refers to the values inferred from the estimated posterior distributions. Results from each model are shown in separate rows with estimates from model (a), (b) and (c) reported in the top, middle and bottom rows, respectively. In the top plots of each row, the points represent the median of the 1,000 samples drawn from the inferred posterior distribution of the merger parameters, whereas the $34^{\rm th}$-$68^{\rm th}$ percentile ranges ($\sigma_{\theta}$) are shown as error bars. The dotted line indicates perfect prediction of the merger parameters.  Mergers from the same galaxy share the same colour.  The distribution of the fractional deviations between the predicted and true parameters compared to $\sigma_{\theta}$ is reported as a sub-panel in each plot. A deviation of $1\,\sigma_{\theta}$ from the true parameter is highlighted as a green-shaded region, while the median distribution of the deviations is show as black continuous line.     By visually inspecting Fig. \ref{fig:cross_validation_cc_galaxy}, it can be noticed that the colour distribution is uniform in all the plots, suggesting that the model performances are similar and do not depend on the assembly history of the test galaxy.

\begin{figure*}
    \centering
    \begin{subfigure}{\textwidth}
        \centering
        \includegraphics[width=0.8\textwidth]{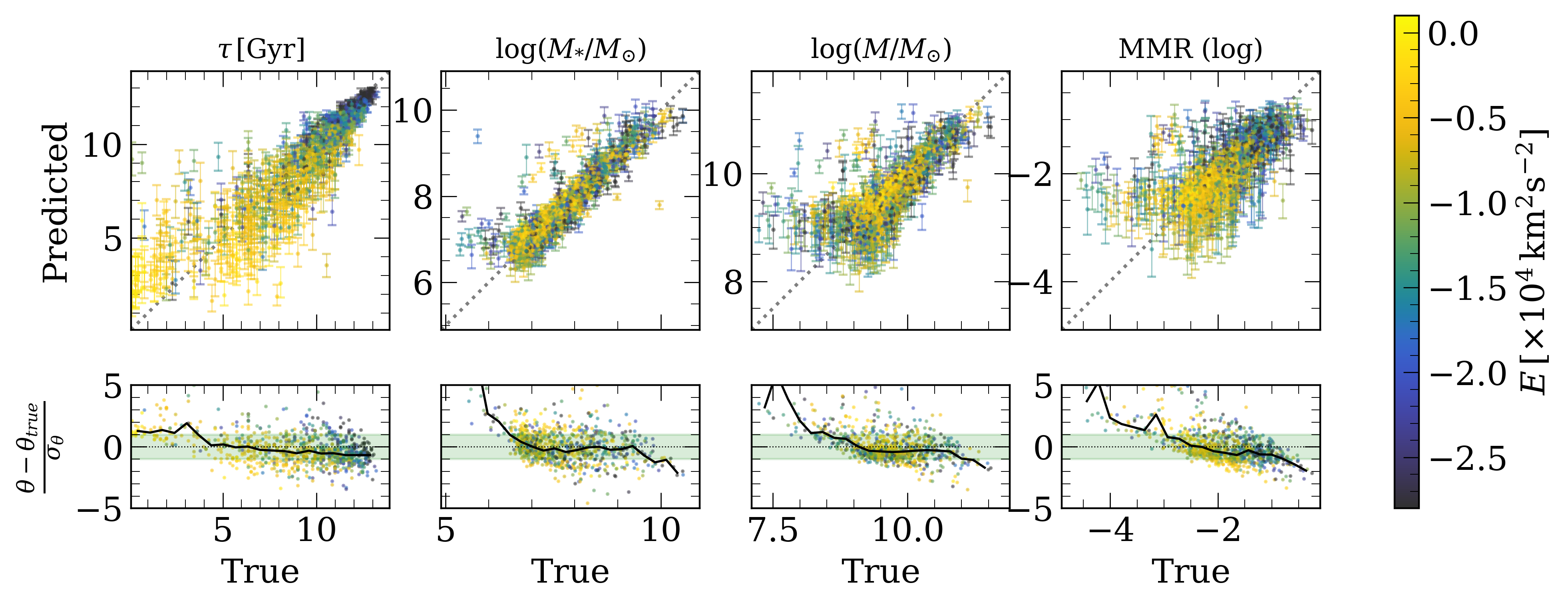}
        \caption{}
        \label{fig:subfig_cc_E}
    \end{subfigure}
    \begin{subfigure}{\textwidth}
        \centering
        \includegraphics[width=0.8\textwidth]{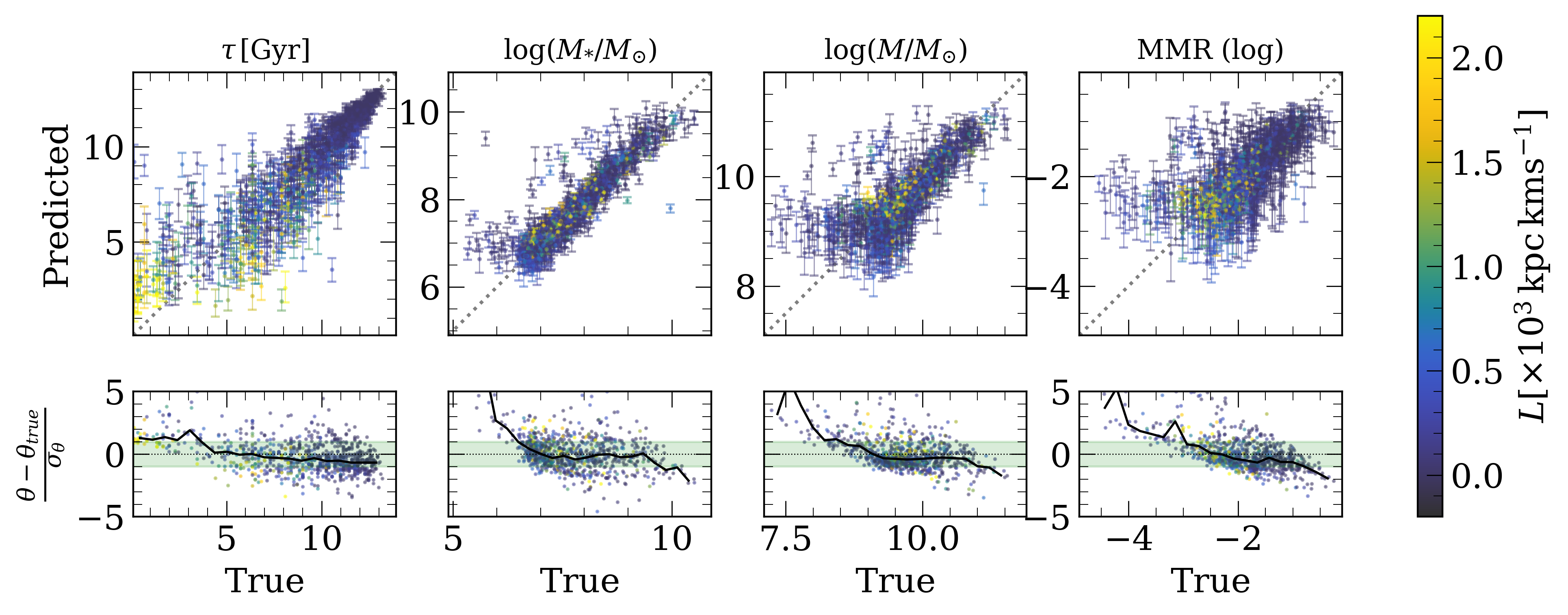}
        \caption{}
        \label{fig:subfig_cc_L}
    \end{subfigure}
    \caption{Same as Fig. \ref{fig:subfig3}, but mergers are colour-coded by the median value of the total energy (a, upper panel) and total angular momentum (b, bottom panel) of the corresponding debris distribution.}
    \label{fig:cross_validation_cc_EL}
\end{figure*}

As intuitively expected from the infall time-binding energy relation of subhalos \citep{Rocha_2012} and conservation principles, the energy and angular momentum distributions of debris appear to contain information on the accretion time of the associated progenitor, while being unrelated to the mass of the system. The energy distribution of debris is strongly dependent on the gravitational potential of the main galaxy; debris from early mergers have had the time to complete a few orbits around the galaxy and sink into the bottom of the gravitational potential, hence having a higher binding energy (i.e. lower total energy) than debris from later mergers. Moreover, as the host halo accretes more mass, the gravitational potential increases overtime contributing to shifting early-accreted stars at lower energies. This is shown in Fig. \ref{fig:subfig_cc_E}, which replicates the plots in Fig. \ref{fig:subfig3} colour-coding the mergers by the median total energy of the associated debris at $z=0$. 

A drop in the performance of model (a) can be noticed for mergers more recent than five billion years ago; rather than being physically motivated, this limitation is likely caused by a lack of training examples in that region as the infall time merger distribution, shown in the relevant sub-panels, is denser at high $\tau$ values. A potential solution could be complementing the current dataset with merger events from other cosmological simulations. Alternatively, a machine learning model could be trained as a surrogate of the simulation process (see e.g. \citealt{Viterbo_CASBI_2024}) to generate an arbitrary number of training examples in any region of the $\bm{\theta}$-parameter space.

The gravitational potential of the main galaxy is closely related to its mass density profile, which is not significantly affected by the accretion of satellites (with the exception of major mergers) as galaxies tend to grow `inside-out' through \textit{smooth accretion} \citep{L'Huillier_massgrowth_2012,Perez_insideoutmassgrowth_2013}. Hence, the energy distribution of debris is not an appropriate proxy for the mass of accreted satellites as shown in Fig. \ref{fig:subfig1}. 
The same conclusion can be drawn from Fig. \ref{fig:cross_validation_cc_EL}, which shows that both the median values of the debris $E$ and $L$ do not correlate with the mass of the associated mergers.

Interestingly, despite failing at predicting exact values, the energy of debris appears to be broadly informative for distinguishing between massive ($\text{MMR} > 1/20$) and less-massive ($\text{MMR} < 1/20$) mergers as shown in Fig. \ref{fig:subfig_cc_E}. When limited to this task, model (a) is able to correctly identify 66 of the 139 massive mergers in the suite, with a 0.48 purity.

\begin{figure*}
    \centering
    \begin{subfigure}{\textwidth}
        \centering
        \includegraphics[width=0.8\textwidth]{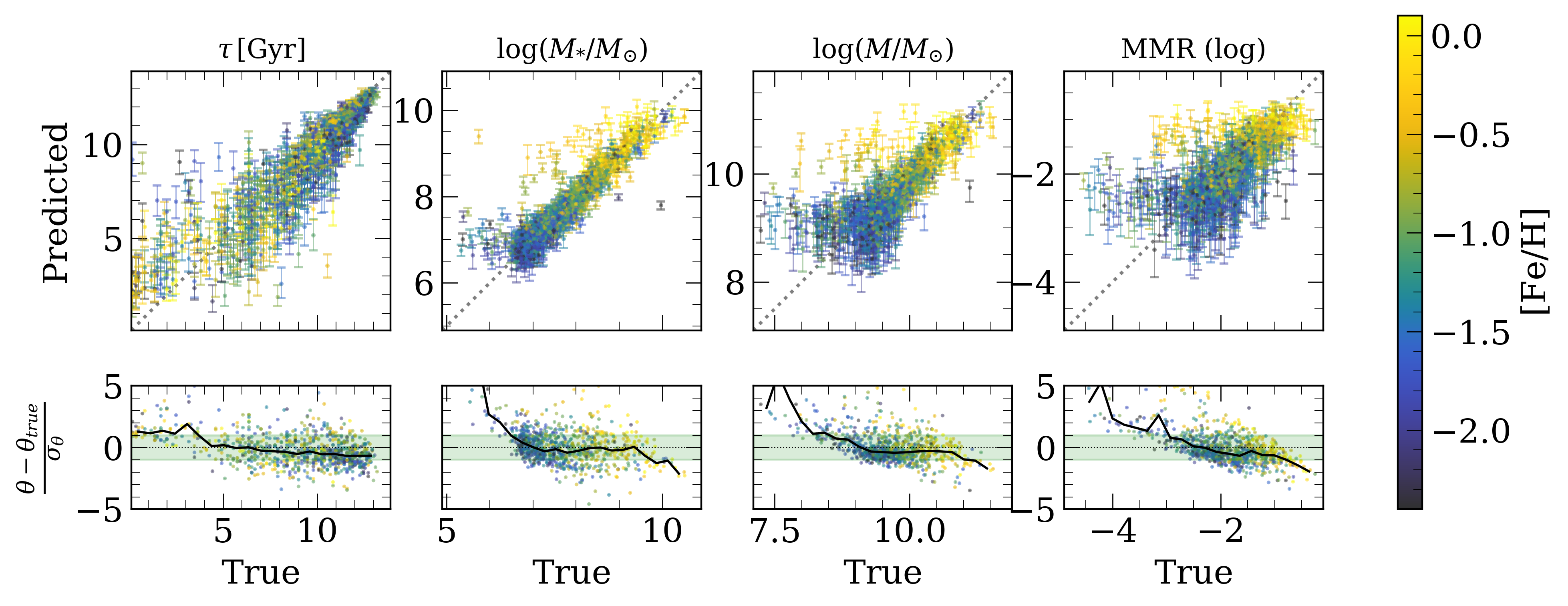}
        \caption{}
        \label{fig:cc_FeH}
    \end{subfigure}
    \begin{subfigure}{\textwidth}
        \centering
        \includegraphics[width=0.8\textwidth]{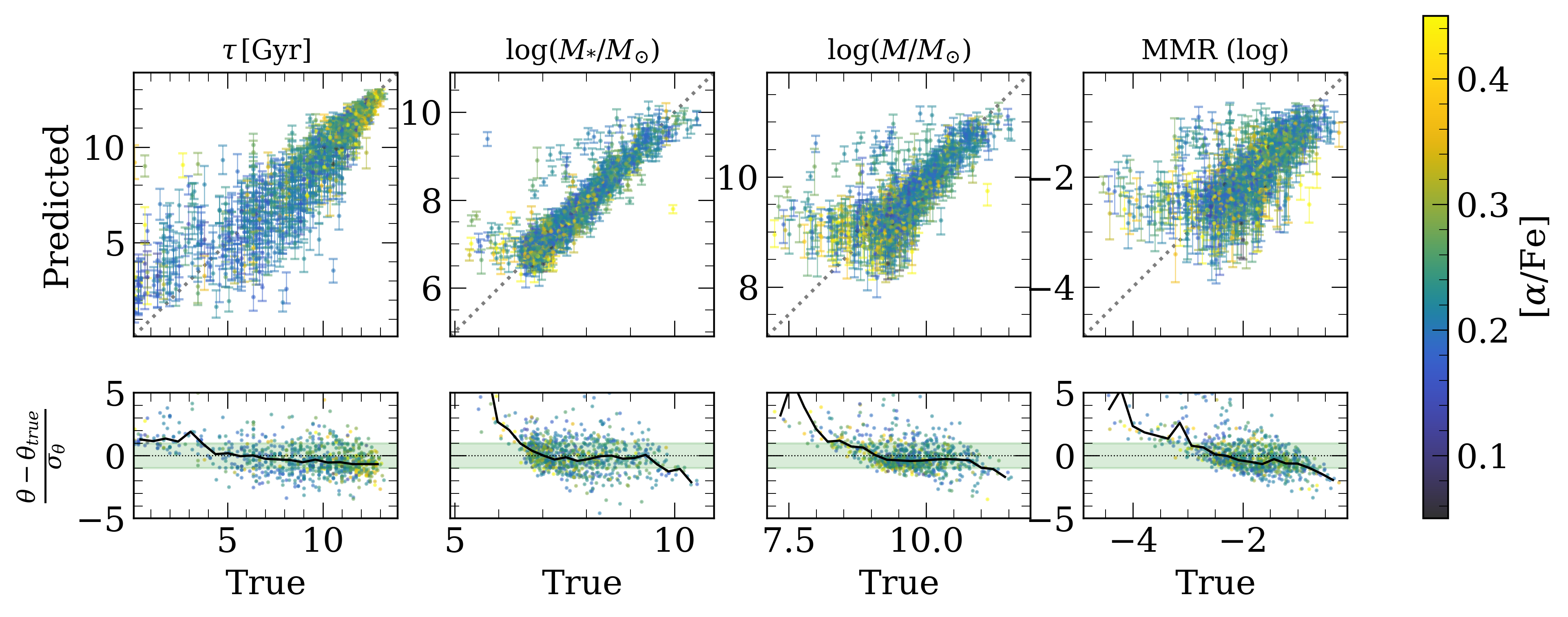}
        \caption{}
        \label{fig:cc_aFe}
    \end{subfigure}
    \caption{Same as Fig. \ref{fig:subfig3}, but mergers are colour-coded by the median value of the $[\text{Fe}/\text{H}]$ (a, upper panel) and $[\alpha/\text{Fe}]$ (b, bottom panel) distributions of the corresponding debris.}
    \label{fig:cross_validation_cc_chem}
\end{figure*}

Providing information on the chemical abundances of debris significantly improves the posterior fitting, as shown by the middle and bottom panels in Fig. \ref{fig:cross_validation_cc_galaxy}. The $[\mathrm{Fe}/\mathrm{H}]$ and $[\alpha/\mathrm{Fe}]$ abundance ratios are closely related to the star formation history of the progenitor galaxy, which is affected both by the accretion time, when it stops due to cold gas being stripped away by the main galaxy, and by the mass of the system, as more massive satellites form multiple generations of stars, hence producing debris with higher metallicity values. Fig.~\ref{fig:cross_validation_cc_chem} shows samples from model (c) trained on the $E$, $L$, $[\mathrm{Fe}/\mathrm{H}]$, and $[\alpha/\mathrm{Fe}]$ debris information and colour-coded by the median values of the actual $[\mathrm{Fe}/\mathrm{H}]$ (Fig. \ref{fig:cc_FeH}) and $[\alpha/\mathrm{Fe}]$ (Fig.~\ref{fig:cc_aFe}) distributions. The $[\mathrm{Fe}/\mathrm{H}]$ abundance of debris correlates well with the mass of the associated progenitor, which is expected from the mass-metallicity relation, as more massive systems formed multiple generations of stellar populations which contributed to enrich the interstellar medium with metals. However, the mass-metallicity relation presents an intrinsic scatter which depends on galaxy-specific properties such as the gas fraction, inflow and outflow, or the star formation rate \citep{vanLoon_MZRscatter_2021}. This could be the cause of the significant deviation between the mass estimates for some metal-rich mergers as can be seen comparing the $\mathrm{log}(M/M_{\sun})$ plots in Figs. \ref{fig:subfig2} and \ref{fig:cc_FeH}. 

Fig.~\ref{fig:cross_validation_cc_chem} suggests that the $[\alpha/\mathrm{Fe}]$ ratio appears to correlate better with the accretion time of mergers than $[\mathrm{Fe}/\mathrm{H}]$. This could be explained in terms of the timescales involved in the two enrichment channels and the effect of the gravitational interaction of the host.  $\alpha$-elements are mostly produced in type-II supernovae on a timescale of few million years, whereas Fe atoms are mostly released during type-Ia supernovae, which require a timescale of  $\sim$ billion year \citep{McWilliam_1997}. When a satellite is accreted, its star formation stops due to the cold gas being stripped away by the host galaxy. Hence,  satellites accreted at late times would appear at a later stage of their chemical evolution with stars characterized by a lower abundance of $\alpha$-elements compared to stars from early mergers \citep{Font_2006}.

The infall time estimates for late mergers ($\tau<5 \, \mathrm{Gyr}$) obtained from models (b) and (c) trained with $[\mathrm{Fe}/\mathrm{H}]$ and $[\alpha/\mathrm{Fe}]$ are closer to the actual values than the ones of model (a) trained only on the $E,\, L$  debris distributions; however, the model accuracy is still not uniform across the whole range of the parameter space. This, alongside with the plateaus at the edges of the predicted $\mathrm{log}(M/M_{\sun})$ and $\mathrm{log}(\mathrm{MMR})$ distributions, is probably related to the limited number of training examples in those ranges. A significant improvement is also seen when distinguishing massive mergers with models (b) and (c) identifying 74 and 88 massive mergers reaching purity values of 0.61 and 0.62, respectively.

A comparison of the performance of the GalactiKit models defined by the three combinations of debris properties can also be performed quantitatively in terms of the root-mean-squared error (RMSE) and mean-relative-uncertainty (MRU). The RMSE is measured for a single merger by directly comparing the inferred values of the parameters $\bm{\theta}$ to the actual $\bm{\theta_{\mathrm{true}}}$ as

\begin{equation}
    \text{RMSE} = \sqrt{\frac{\sum_{i=1}^{N}(\bm{\theta_i}-\bm{\theta_{\mathrm{true}}})^{2} }{N}},
\end{equation}
where $N=1,000$ is the number of $i$-samples drawn from the approximated posterior distribution of each merger event. The RMSE estimates the accuracy of the model by measuring how close the average prediction of the inferred merger parameters is to the true value. 

The MRU is defined comparing the size of a given confidence interval, defined by a percentile range $Q_{a}(\bm{\theta}) -  Q_{b}(\bm{\theta})$, to the extent of the prior distributions of each dimension of $\bm{\theta}$. Each prior distribution informs on the probability that a merger property can assume a certain value, hence excluding the ones that are not physical. In this study, the four merger parameters that define the properties of the progenitors at infall are assumed to have uniform prior distributions such that, for a given merger, the true parameters are known to have equal probability to lie within any given point of the corresponding ranges, i.e. $\tau \,/\,\mathrm{Gyr} \in \{0,13.5\}$, $\log (M_{*}/M_{\sun}) \in \{5,11\}$, $\log (M/M_{\sun}) \in \{7,12\}$ and $\log (\mathrm{MMR}) \in \{-5,0\}$. Thus, the MRU for the predictions of the properties of a given merger event is calculated as 

\begin{equation}
    \text{MRU}(a,b) = \left| \frac{Q_{a}(\bm{\theta}) -  Q_{b}(\bm{\theta})}{\mathrm{max}(\bm{\theta_\mathrm{true}})-\mathrm{min}(\bm{\theta_\mathrm{true}})} \right |,
\end{equation}
where $Q_a(\bm{\theta})$ and $Q_b(\bm{\theta})$ are the values of $\bm{\theta}$ at the $a^{\rm th}$ and $b^{\rm th}$  percentile of the predicted posterior distribution, and $\mathrm{max}(\bm{\theta_\mathrm{true}})-\mathrm{min}(\bm{\theta_\mathrm{true}})$ refers to the difference between the maximum and minimum values of $\bm{\theta}$ as defined by the prior ranges. The MRU provides an estimate of the precision of  the predictions, where MRU values approaching $0$ indicate a model that consistently makes similar predictions for a given merger (i.e. the model is confident about the estimated value of $\bm{\theta}$), whereas MRU values close to $1$ suggest that, despite the information of the debris properties, the model is not able to constrain the merger parameters to a specific region of the prior space.  

The RMSE and MRU were computed for each of the 928 merger events considering the $\bm{\theta}$ samples used for producing Figs. \ref{fig:cross_validation_cc_galaxy}, \ref{fig:cross_validation_cc_EL} and \ref{fig:cross_validation_cc_chem}. The average of the RMSE and MRU values are reported in Table~\ref{tab:stats_total}, alongside with the average fraction of test examples for which the true merger parameters fall within the $34^{\rm th}$ - $68^{\rm th}$ and $5^{\rm th}- 95^{\rm th}$ percentile ranges of the predicted posterior distributions.

\begin{table*}
    \centering
    \caption{Quantitative comparison between the merger parameter estimates inferred through the posterior models trained on different combinations of debris information. From left to right, the columns report: i) the inferred merger parameter; ii) the root-mean-squared error (RMSE) between the actual and inferred merger parameter; iii) the fraction of mergers whose actual parameters are within the $34^{\rm th}-68^{\rm th}$ percentile range of the samples drawn from the posterior model; iv) the mean-relative uncertainty (MRU) for the $34^{\rm th}-68^{\rm th}$ percentile range; v) and vi) are the same as iv) and v) but considering the $5^{\rm th}-95^{\rm th}$ percentile range as confidence interval.}
    \begin{tabular}{|c c c c c c|}
    \hline 
    $\theta$ & RMSE & within $34^{\rm th} - 68^{\rm th}$ & MRU(34,68) & within $5^{\rm th}-95^{\rm th}$ & MRU(5,95)\\ 
    \hline
    \multicolumn{6}{| c |}{$E$, $L$} \\
    \hline
    $\tau$ &  3.44 & 0.28 & 0.15 & 0.81 &  0.56\\
    $\mathrm{log}(M_{*}/M_{\sun})$ & 1.25 & 0.28 & 0.14 & 0.88 & 0.47\\
    $\mathrm{log}(M/M_{\sun})$ & 1.00 & 0.36 & 0.14 & 0.91 & 0.51\\
    $ \mathrm{MMR} (\mathrm{log}) $ & 0.94 & 0.39 & 0.12 & 0.91 & 0.48\\
    \hline
    \multicolumn{6}{| c |}{$[\mathrm{Fe}/\mathrm{H}]$, $[\alpha/\mathrm{Fe}]$} \\
    \hline
    $\tau$ & 1.92 & 0.30 & 0.08 & 0.88 &  0.33\\
    $\mathrm{log}(M_{*}/M_{\sun})$ & 0.58 & 0.39 & 0.05 & 0.92 & 0.24\\
    $\mathrm{log}(M/M_{\sun})$ & 0.73 & 0.41 & 0.08 & 0.92 & 0.38\\
    $ \mathrm{MMR} (\mathrm{log}) $ & 0.87 & 0.38 & 0.10 & 0.91 & 0.46\\
    \hline
    \multicolumn{6}{| c |}{$E$, $L$, $[\mathrm{Fe}/\mathrm{H}]$, $[\alpha/\mathrm{Fe}]$} \\
    \hline
    $\tau$ & 1.83 & 0.29 & 0.08 & 0.86 &  0.30\\
    $\mathrm{log}(M_{*}/M_{\sun})$ & 0.54 & 0.35 & 0.04 & 0.91 & 0.22\\
    $\mathrm{log}(M/M_{\sun})$ & 0.69 & 0.34 & 0.08 & 0.90 & 0.35\\
    $ \mathrm{MMR} (\mathrm{log}) $ & 0.81 & 0.34 & 0.10 & 0.89 & 0.41\\
    \hline
    \end{tabular}
    \label{tab:stats_total}
\end{table*}

Compared to model (a) informed only through dynamical data, the GalactiKit models (b) and (c) trained including $[\mathrm{Fe}/\mathrm{H}]$ and $[\alpha/\mathrm{Fe}]$ provide the most accurate predictions of the merger parameters with an average decrease in RMSE of about 45, 55, 29 and 12 per cent for the estimates of $\tau, \, \mathrm{log}(M_{*}/M_{\sun}), \, \mathrm{log}(M/M_{\sun}) \, \mathrm{and} \, \mathrm{log}(\text{MMR})$, respectively. As previously discussed, the time estimates of all models are considerably more accurate for merger events occurred earlier than 5 billion years ago with average RMSE values of 3.30, 1.75 and 1.64 Gyr compared to 4.45, 3.12 and 3.09 Gyr for late mergers as predicted by models (a), (b) and (c), respectively. This would be less problematic for Galactic archaeology studies as significant satellites of the Milky Way with $\tau<5\,\mathrm{Gyr}$ are expected to be either still orbiting the Galaxy or traceable from stellar streams.

Estimates for the stellar and halo masses of mergers are also more accurate when provided by the models (b) and (c). In both cases, the stellar mass is better predicted than the halo mass because of the intrinsic scatter in the stellar mass-halo mass relation (SMHR), which could be due, for instance, to halos of same total mass but different concentration forming a different number of stars as more concentrated halos are associated to earlier formation times \citep{Matthee_scatterSMHR_2017}. The scatter in the SMHR is found to be larger at smaller halo masses, which could also be the cause to the decrease in accuracy for the halo mass estimates of mergers with $\mathrm{log}(M/M_{\sun})<9$.

Table \ref{tab:stats_total} indicates that all models predict less than half of the merger parameters within the $34^{\rm th}-68^{\rm th}$ percentile range; as the RMSE are reasonably low, this is a sign of overconfidence probably associated to both the reduced variety of merging events in the training set and the repetition of samples from the same progenitors. The small size of the confidence interval compared to the parameter space, as reported by the MRU values, points towards the same conclusion. More conservative estimates of the merging parameters could be obtained taking into account a larger confidence interval, as shown in the last two columns of Table \ref{tab:stats_total}; however, there is the risk of obtaining an uninformative analysis as the size of the considered percentile range approaches the extent of the prior ranges of the parameters.

Overall, model (b) trained exclusively on $[\mathrm{Fe}/\mathrm{H}]$ and $[\alpha/\mathrm{Fe}]$ has a very similar performance compared to model (c) trained including $E$ and $L$. Thus, the information contained in the chemical abundance ratios of debris appears to be enough for inferring the infall time and mass of the associated merger. The extra information provided by the dynamical quantities seems to contribute to the overconfidence of the model as the decrease in MRU values comes with a slight decrease in the fraction of parameters correctly inferred for both the 34th-68th and 5th-95th percentile ranges. However, the information on the $E$ and $L$ distributions of debris also appears to help constraining the mass predictions of metal-rich, but relatively low-mass progenitor galaxies, as can be noticed comparing the $\log(M_*/M_{\sun})$, $\log(M/M_{\sun})$ and $\log(\mathrm{MMR})$ plots in Figs. \ref{fig:subfig2}, \ref{fig:subfig3}, and \ref{fig:cc_FeH}. This is also reflected with a difference in RMSE (0.54 and 0.58 dex; 0.69 and 0.73 dex; 0.81 and 0.87 dex) between model (c) and (b) as shown in Table \ref{tab:stats_total}. Hence, the combination of chemical and dynamical debris properties appears to be preferred for the development of models of the merger parameters, if the observed overconfidence is addressed through changes in the data, e.g.,  increasing the variety of training examples, or methodology, e.g., changing the technique for the posterior estimation.

\section{Discussion}
\label{sec:discussions}

GalactiKit provides a quantitative description of the assembly history of Milky Way-like galaxies by returning predictions within confidence intervals of the infall time, stellar mass, halo mass, and mass ratio with the host of their progenitors. The precision of the prediction depends on the amount of information available on the distribution of the merger debris $z=0$, with models trained on chemical abundances significantly outperforming those based only on dynamical data. Complementing chemical with dynamical information seems to mostly contribute to reducing the scatter in the mass estimates of metal-rich mergers, while maintaining a similar performance to the models relying on chemical abundances only in the rest of cases. 

Hence, there appears to be a strong link between the properties of a galaxy at infall and the chemical abundance distribution of its debris at $z=0$. This can be motivated by known physical phenomena such as the observed relation between mass and metallicity of galaxies \citep{Harmsen_massmetallicity_2017} and the expected difference in $\alpha$-element abundances expected in late and early accreted satellites \citep{Font_2006,Grimozzi2024}. However, the GalactiKit models leverage the knowledge of these galaxy formation processes encoded in the Auriga chemical evolution model, which is limited by theoretical uncertainties in stellar evolution such as the nucleosynthetic yields of the different enrichment channels or the characteristics of the intermediate mass function. Therefore, potential drops in the predictive ability of the models could be observed if applied to a cosmological context where the chemical evolution of galaxies significantly differs from the one described in Auriga. Testing on a suite of simulations with different chemical evolution prescription could offer a framework to evaluate the potential drop in performance before the application to observations. 

Another limitation of the GalactiKit methodology is the assumption that accreted stars can be grouped into their progenitors of origin even in galaxies whose merger tree information is not accessible. Based on the idea that stars which formed in the same system share similar orbital and chemical properties, several methods have been developed to identify accreted substructures in the Milky Way with clustering algorithms \citep[e.g.,][]{Koppelman_2019,Naidu_substructures_2020, Dodd_catalogue_2023}. However, the purity of the resulting merger debris groups is unknown and can be affected by the significant contamination of in-situ stars. Although there are methods to mitigate the presence of the in-situ population \citep{Sante_2024}, how well the GalactiKit models can infer the properties of disrupted progenitors based on incomplete or contaminated merger debris distributions remains to be investigated.

In spite of the limitations mentioned above, the GalactiKit methodology is promising for future applications to the Milky Way. Combined with the information on the merger debris distribution of the accreted substructures already discovered in the Milky Way, GalactiKit can be used to provide a quantitative picture of the formation history of the Galaxy by estimating the mass and infall time of the various accretion events. A similar endeavour was attempted by \cite{Kruijssen_kraken_2019}, who used the statistical correlations between globular cluster and galaxy properties in the E-MOSAICS simulations \citep{Kruijssen_emosaics_2019} to reconstruct the merger tree of the Milky Way from the observed globular cluster population. GalactiKit provides a framework to extend this analysis from the globular cluster population to the millions of single stars observed by Gaia and to the present and upcoming spectroscopic surveys, hence depicting a more complete picture of the assembly history of the Milky Way.

A detailed reconstruction of the turbulent past of the Galaxy could be useful for a better understanding of the evolution of the Milky Way in a cosmological context. The Galaxy is considered to have undergone an unusually quiet accretion history within the \textLambda CDM cosmology framework \citep{Hammer_2007,Tilly_2020}. Theoretical predictions from the dark matter mass function \citep{Tinker_halomassfn_2008} and the stellar-to-halo mass relation \citep{Brook_smhm_2014} suggest that the majority of merger events for a Milky Way-mass galaxy consists of the accretion of low-mass dark halos \citep{Purcell_2007}, with the majority of stars in the stellar halo coming from only $\sim$3 major (mass ratio $> 1/3$), luminous satellites \citep{Fakhouri_mergerrates_2010}. In the case of the Galaxy, merging events in this mass range correspond to the accretion of the GS/E dwarf galaxy and the Large Magellanic Cloud, which are more than 10 billion years apart. Adding further complexity, \cite{Hammer_2024} showed that the accretion time of the GS/E dwarf galaxy, as measured using the age-metallicity relation (AMR) of the globular cluster population identified by \citet{Kruijssen_kraken_2019}, would have had to be shifted at even earlier times to be consistent with the satellite population of the Milky Way under the infall time-binding energy relation \citep{Rocha_2012}.
Thus, the application of GalactiKit models could provide a clearer perspective by estimating the accretion times of all the disrupted satellites of the Galaxy, including those without an associated globular cluster population.

Studying the properties of the progenitors of the Milky Way can also be useful outside the Galactic archaeology context. \cite{Limberg_2024} suggested that the evidence for an intermediate-mass black hole (IMBH) in \textomega Centauri \citep{Haberle_IMBH_2024}, arguably the old nuclear star cluster of the GS/E galaxy \citep{Limberg_2022}, could extend the central black hole mass-host stellar mass relation \citep{Greene_IMBHscalingrelations_2020} to the dwarf galaxies regime. Hence, more IMBHs could potentially be located within the proximity of debris from disrupted progenitors of the Milky Way. The presence of BHs of extragalactic origin has already been confirmed by \cite{Balbinot_2024} who associated the $33 \, M_{\sun}$ BH found by \cite{Gaia_BH_2024} to the ED-2 stream \citep{Dodd_catalogue_2023}, believed to be the remnant of a star cluster $10^{3} - 10^{4} \, M_{\sun}$ in mass. Therefore, GalactiKit could be used to estimate the mass of the disrupted progenitors of the Milky Way directly from the distributions of the merger debris, hence informing on the potential presence of undetected IMBHs among the accreted stars.




\section{Conclusions}
\label{sec:conclusions}

GalactiKit is a data-driven methodology for determining the lookback infall time, stellar mass, halo mass and mass ratio with the host of disrupted satellite galaxies of Milky Way analogues. Using simulation-based inference to leverage the information on galaxy formation encoded in the merger trees of the Milky Way-mass halos from the Auriga cosmological simulations, GalactiKit estimates the posterior distribution of the merger parameters as conditioned by the $z=0$ distribution of the corresponding debris.   

In order to investigate how accurate is the inference of merger properties from various degrees of information on the merger debris, we developed three GalactiKit models considering the following chemo-dynamical properties of the accreted stars: (a) $E$ and $\,L$, (b) $[\text{Fe}/\text{H}]$ and $[\alpha/\text{Fe}]$ and (c) $E, \, L, \, [\text{Fe}/\text{H}]$ and $[\alpha/\text{Fe}]$. The results of our analysis can be summarised as follows:

\begin{enumerate}
    \item the total energy of accreted stars is a tracer of the accretion time of progenitor galaxies (Fig. \ref{fig:subfig1}); this result has already been known for surviving satellites \citep{Rocha_2012} and is extended to phase-mixed substructures here. Moreover, energy and angular momentum information can be used for an approximate identification of massive mergers; model (a) correctly predicts a median MMR above 1/20 for 47 per cent of the total massive mergers, with a purity of 0.48.
    \item including the chemical information of the debris provides a significant improvement in the accuracy of the estimates for all merger parameters, and is a necessary ingredient for the prediction of the stellar and halo masses of progenitor galaxies at infall (Fig. \ref{fig:cross_validation_cc_galaxy}). For both models (b) and (c), stellar masses are better predicted than halo masses. This is probably related to the intrinsic scatter in the stellar mass-halo mass relation which depends on the chemical and stellar evolution specific to each progenitor galaxy. 
    \item model (c), informed by both the chemical and dynamical properties of the merger debris, is the most accurate at the estimation of the merger parameters with an average RMSE of $1.83\,\mathrm{Gyr}$, $0.54\,\mathrm{dex}$, $0.69\,\mathrm{dex}$ and $0.81 \, \mathrm{dex}$ for $\tau/[\mathrm{Gyr}]$, $\mathrm{log}(M_{*}/M_{\sun})$,  $\mathrm{log}(M/M_{\sun})$ and $\mathrm{log}(\mathrm{MMR})$, respectively. Interestingly, as shown in Table \ref{tab:stats_total}, predictions of similar accuracy can be achieved with $\mathrm{model}\, (\mathrm{b})$ trained exclusively on the chemical abundance ratios distribution of debris, which is also found to provide slightly more conservative estimates. However, as can be seen comparing the $\log(M_*/M_{\sun})$, $\log(M/M_{\sun})$ and $\log(\mathrm{MMR})$ plots in Figs. \ref{fig:subfig2}, \ref{fig:subfig3}, and \ref{fig:cc_FeH}, $\mathrm{model}\, (\mathrm{b})$ tends to over-estimate the mass of high-metallicity and relatively low-mass progenitor galaxies, which is moderated in model (c) by the addition of dynamical information.
    \item  all models show a large scatter in the parameter estimates for mergers that are not well represented in the training set. This is due to the limited number of merger events in the simulations which was not successfully addressed in the data pre-processing step. Potential solutions involve combining data from multiple cosmological simulations, or implementing a surrogate model for data generation. A significant difference in the accuracy is observed between the infall time predictions of early ($\tau>5\, \mathrm{Gyr}$) and late ($\tau<5\, \mathrm{Gyr}$) mergers as more accurate predictions are made for the former. 
\end{enumerate}

In conclusion, we have shown that GalactiKit is an efficient method in the study of the assembly history of Milky Way-like systems. This methodology can be efficiently applied to the analysis of the chemo-dynamical data of millions of stars observed by current and upcoming surveys in the Milky Way, to better constrain the infall times and the masses of disrupted progenitors.

\section*{Acknowledgements}

AS acknowledges a Science Technologies Facilities Council (STFC) PhD studentship at the LIV.INNO Centre for Doctoral Training ``Innovation in Data Intensive Science'' and funding from The Alan Turing Institute through the ``Enrichment Scheme Award''. This study made use of Prospero high-performance computing facility at Liverpool John Moores University. This work was also partly supported by the UK's Science \& Technology Facilities Council (STFC grant ST/S000216/1, ST/W001136/1). RJJG is supported by an STFC Ernest Rutherford Fellowship (ST/W003643/1). DK acknowledges MWGaiaDN, a Horizon Europe Marie Sk\l{}odowska-Curie Actions Doctoral Network funded under grant agreement no. 101072454 and also funded by UK Research and Innovation (EP/X031756/1). 

\section*{Data Availability}

The Auriga simulations can be downloaded through the Globus file transfer service (\url{https://globus.org/}). Detailed instructions on how to access the data are provided in \url{https://wwwmpa.mpa-garching.mpg.de/auriga/data.html}. 

The codes and models developed in the analysis are available under reasonable request to the authors.



\bibliographystyle{mnras}
\bibliography{references} 



\bsp	
\label{lastpage}
\end{document}